\documentclass[aps,prd,preprint,twoside,showpacs,superscriptaddress,nofootinbib,amsmath,amssymb]{revtex4-1}

\usepackage{MnSymbol}
\usepackage[utf8]{inputenc}
\usepackage{soulutf8}
\usepackage{graphicx}
\usepackage{epstopdf}
\usepackage{amsmath}
\usepackage{slashed}
\usepackage{xcolor}
\usepackage{mathtools}
\usepackage{array,multirow}
\usepackage{booktabs}
\usepackage{adjustbox} 
\usepackage{rotating}
\usepackage{blindtext}
\usepackage{physics}
\usepackage{caption}
\usepackage{subcaption}
\usepackage{hyperref}
\hypersetup{colorlinks=true,linkcolor=purple,anchorcolor=blue,citecolor=blue, filecolor=blue,urlcolor=red,bookmarksnumbered=true,
	pdfview=FitB
}

\begin{document}



\title{Gluon Distributions in the Pion}

\author{Satvir Kaur}
\email{satvir@impcas.ac.cn}

\affiliation{\small Institute of Modern Physics, Chinese Academy of Sciences, Lanzhou 730000, China}
\affiliation{\small School of Nuclear Science and Technology, University of Chinese Academy of Sciences, Beijing 100049, China}
\affiliation{\small CAS Key Laboratory of High Precision Nuclear Spectroscopy, Institute of Modern Physics, Chinese Academy of Sciences, Lanzhou 730000, China}

\author{Chandan Mondal}
\email{mondal@impcas.ac.cn}

\affiliation{\small Institute of Modern Physics, Chinese Academy of Sciences, Lanzhou 730000, China}
\affiliation{\small School of Nuclear Science and Technology, University of Chinese Academy of Sciences, Beijing 100049, China}
\affiliation{\small CAS Key Laboratory of High Precision Nuclear Spectroscopy, Institute of Modern Physics, Chinese Academy of Sciences, Lanzhou 730000, China}

\begin{abstract} 
We formulate a light-front model for the pion that explicitly incorporates the gluonic degree of freedom. In this framework, high-energy scattering off the pion is described by an active gluon, while the remaining constituents are treated as a spectator system with an effective mass. The mass spectrum and light-front wave functions (LFWFs) of the pion are determined by solving two Schr\"odinger-like equations derived from quantum chromodynamics: the light-front holographic equation in the chiral limit and the 't~Hooft equation. Using the resulting LFWFs, we compute the gluon distributions within pion, including the parton distribution functions in comparison with the available global fits, generalized parton distributions, and transverse momentum-dependent distributions. 
The present analysis is restricted to the Dokshitzer–Gribov–Lipatov–Altarelli–Parisi (DGLAP) domain of the pion’s gluon GPD, without explicit construction of the Efremov–Radyushkin–Brodsky–Lepage (ERBL) region. Consequently, the polynomiality condition for full Mellin moments is not satisfied by construction.
Furthermore, we demonstrate that the obtained LFWFs provide a good description of the gravitational form factors of pion when compared with the recent lattice QCD results.

\end{abstract}

\maketitle
\section{Introduction}
Quantum Chromodynamics (QCD) is the fundamental theory of the strong force~\cite{Callan:1977gz}, describing hadrons as bound states of quarks and gluons. Despite its successes, key aspects of hadron structure—particularly the role of gluons—remain poorly understood. Gluons are essential for binding quarks and generating mass, yet their distribution inside light mesons such as the pion is still not well-determined.

The pion plays a unique role in QCD: it is the lightest meson, a quark-antiquark bound state, and the Nambu-Goldstone boson of spontaneously broken chiral symmetry~\cite{Nambu:1961tp,Nambu:1961fr}. Its simplicity makes it an ideal system for probing nonperturbative QCD. While most theoretical studies have focused on valence quark structure or incorporated gluons only perturbatively~\cite{deMelo:2008rj,Frederico:2009fk,Chang:2013pq,Pasquini:2014ppa,Gutsche:2014zua,Lorce:2016ugb,Bacchetta:2017vzh,deTeramond:2018ecg,Ahmady:2018muv,Ahmady:2020mht,Ahmady:2019yvo,Raya:2021zrz,Lan:2019vui,dePaula:2022pcb,Lu:2022cjx,Cui:2022bxn,Mondal:2021czk,Adhikari:2021jrh,Ahmady:2021lsh,Ahmady:2021yzh,Lan:2020fno}, recent advances in lattice QCD~\cite{Fan:2021bcr,Shanahan:2018pib} and phenomenological models~\cite{Pasquini:2023aaf,Lan:2021wok,Zhu:2023lst,Lan:2024ais,Broniowski:2022iip} now explicitly account for gluon dynamics, underscoring their importance in pion structure.

Light-front quantization offers a powerful framework for studying hadrons through light-front wave functions (LFWFs)~\cite{Brodsky:1997de}, which encode partonic configurations and enable the calculation of hadronic matrix elements for both inclusive and exclusive processes~\cite{Lorce:2011dv}. This approach is well suited for investigating the partonic structure of hadrons.

In this work, we compute gluon-related observables in the pion including the parton distribution functions (PDFs), generalized parton distributions (GPDs), and transverse momentum dependent distributions (TMDs). 
PDFs describe the longitudinal momentum distribution of partons, while GPDs extend this framework to include spatial and momentum correlations. 
They depend on the longitudinal momentum fraction $x$, the skewness $\xi = -\Delta^+ / 2\bar{P}^+$, and the squared momentum transfer $t = \Delta^2=-Q^2<0$, 
where $\bar{P} = (P' + P)/2$ is the average pion momentum and $\Delta = P' - P$ is the momentum transfer.
Their moments relate to gravitational form factors (GFFs), revealing insights into hadron mass, pressure, and energy distributions~\cite{Ji:1996ek,Polyakov:2018zvc,Lorce:2018egm,Burkert:2023wzr}. TMDs further enrich this picture by resolving transverse parton momenta.

Global analyses of pion PDFs, primarily using Drell-Yan, $J/\psi$ production, and direct photon data~\cite{Barry:2021osv,Novikov:2020snp,Pasquini:2023aaf, Kotz:2025lio}, have advanced our understanding of its structure. However, the gluon and sea-quark content remains less constrained than in the nucleon due to limited experimental data. Upcoming facilities like Jefferson Lab~\cite{Arrington:2021alx} and future electron-ion colliders (EICs)~\cite{AbdulKhalek:2021gbh,Anderle:2021wcy} aim to probe pion PDFs via the Sullivan process~\cite{Sullivan:1971kd}.
The COMPASS++/AMBER program will further explore pion structure using high-intensity pion beams~\cite{Adams:2018pwt}.

Experimental access to GPDs is challenging but progressing through deeply virtual Compton scattering (DVCS) and vector meson production~\cite{Diehl:2023nmm,H1:1999pji,HERMES:2001bob}, with data from H1, ZEUS, and HERMES~\cite{H1:2001nez,H1:2005gdw,H1:2009wnw,ZEUS:2003pwh,HERMES:2010nas,HERMES:2012gbh,HERMES:2012idp} already providing constraints. Near-threshold photo- and leptoproduction of charmonium and bottomonium provides a powerful probe of the hadron’s gluonic structure~\cite{Mamo:2019mka,Mamo:2022eui,Guo:2023qgu,Hatta:2018ina,Boussarie:2020vmu,Pentchev:2024sho,GlueX:2019mkq,GlueX:2023pev}. Future EICs are expected to extend these studies~\cite{AbdulKhalek:2021gbh,Anderle:2021wcy}, while Jefferson Lab has already reported promising results from near-threshold $J/\psi$ production~\cite{GlueX:2019mkq,Duran:2022xag,GlueX:2023pev,Liu:2024yqa,Hechenberger:2024abg}.

While TMDs have been extensively studied for the proton via SIDIS, DY, and $e^+e^-$ annihilation~\cite{Diehl:2023nmm}, pion TMDs remain less explored, with only limited extractions from DY data~\cite{Wang:2017zym,Vladimirov:2019bfa,Cerutti:2022lmb}. Theoretical investigations of pion TMDs have been pursued in Refs.~\cite{Pasquini:2014ppa,Noguera:2015iia,Lorce:2016ugb,Bacchetta:2017vzh,Ceccopieri:2018nop,Ahmady:2019yvo,Shi:2020pqe,Kaur:2020vkq}, but further experimental and theoretical efforts are required.

Recent progress on gluon structure has largely focused on the nucleon. 
Active-gluon spectator approaches have been employed to investigate gluon PDFs and TMDs~\cite{Chakrabarti:2023djs, Lyubovitskij:2021qza, Lyubovitskij:2020xqj}, extended to GPDs at nonzero skewness in the DGLAP region~\cite{Chakrabarti:2024hwx}, and more recently to gluon contributions to GFFs~\cite{Sain:2025kup} and GTMDs~\cite{Chakrabarti:2025qba}. 
In parallel, Hamiltonian-based methods such as Basis Light-Front Quantization (BLFQ)~\cite{Zhang:2025nll, Lin:2024ijo, Kaur:2024iwn, Lan:2021wok, Lin:2023ezw}, and Bethe–Salpeter/Dyson-Schwinger equations approaches~\cite{Lan:2024ais,Hutauruk:2022zju,Freese:2021zne} dynamically generate higher Fock components and have been applied to gluon distributions in the proton, pion, and spin-1 mesons. Recent string-based holographic QCD work also treats nucleon GPDs in both the DGLAP and ERBL regions~\cite{Mamo:2024jwp, Mamo:2024vjh, Hechenberger:2025wnz}.
Complementary developments within the holographic QCD framework have further addressed gluon  PDFs, GFFs, and  GPDs for both the proton and the pion~\cite{ deTeramond:2021lxc,Gurjar:2022jkx}. 

In this work, we develop a light-front model of the pion that explicitly incorporates gluonic degrees of freedom, providing a framework to explore the pion’s gluonic structure. In this approach, high-energy scattering processes involving the pion are modeled as being mediated by an active gluon, while the remaining constituents form an effective spectator system with spin one and an effective mass, motivated by the gluon-spectator model for the nucleon~\cite{Lyubovitskij:2021qza,Chakrabarti:2023djs}.
The mass and LFWFs of the pion are determined self-consistently by solving two complementary equations rooted in QCD: the light-front holographic Schr\"odinger equation, which captures the transverse dynamics and confinement mechanism in the chiral
limit~\cite{Brodsky:2014yha}, and the ’t Hooft equation, which governs the longitudinal momentum structure of mesons in the large-$N_c$ limit~\cite{tHooft:1974pnl}. Together, these equations offer a three-dimensional description of
the pion’s bound-state dynamics, incorporating both the transverse and longitudinal aspects
of confinement.
This unified framework provides predictions for gluon-sensitive observables, including the gluon PDFs, GPDs, TMDs, and GFFs of the pion. The model predictions show good agreement with recent lattice QCD results and the available global fits, demonstrating its effectiveness in capturing both the momentum and spatial structure of the pion.

\section{Theoretical framework}

Holographic light-front QCD (hLFQCD) is formulated in the chiral limit of light-front QCD and establishes a correspondence between strongly coupled (3+1)-dimensional light-front QCD and weakly coupled string modes in a (4+1)-dimensional anti-de Sitter (AdS) space. For a comprehensive review of hLFQCD, see Ref.~\cite{Brodsky:2014yha}.
A central prediction of hLFQCD is that the pion, as the lightest bound state, becomes massless in the chiral limit. Moreover, the model predicts that meson masses lie on universal Regge trajectories, in agreement with experimental observations. The slopes of these trajectories are governed by the strength of the confining potential, characterized by the parameter $\kappa$. In this framework, the confining potential in physical spacetime emerges from a dilaton field that breaks the conformal symmetry of AdS space. Specifically, a quadratic dilaton profile in the fifth dimension leads to a light-front harmonic oscillator potential, which has proven to be phenomenologically successful.
The scale parameter $\kappa$ is fixed by fitting the experimentally measured slopes of Regge trajectories across various meson families, yielding a universal value of $\kappa \simeq 0.5$ GeV for light mesons~\cite{Brodsky:2014yha}.

The minimal Fock state of a meson consists exclusively of a valence quark-antiquark pair. When higher Fock sectors are included, the meson wavefunction acquires additional components involving gluons and sea quark-antiquark pairs. In our simplified model, we describe the meson as a composite system made up of an active gluon (mass $m_1$ and spin $\lambda_g$) and an effective spin-1 spectator (mass $m_2$ and spin $\lambda_s$).
We work in standard light-front coordinates, where $v^\pm = \frac{1}{\sqrt{2}}(v^0 \pm v^3)$ and $v_\perp = (v^1,v^2)$. The longitudinal momentum fraction of the active gluon is denoted by $x = k^+/P^+$, with $k^+$ and $P^+$ the light-front momenta of the gluon and the pion, respectively. The transverse separation between the gluon and the spectator is represented by $\vec{b}_\perp = b_\perp e^{i\varphi}$.
The LFWF $\Psi(x,\vec{b}_\perp)$ provides a Lorentz-invariant description of the gluon-spectator dynamics in terms of these light-front variables.
To establish a connection with AdS space, the holographic variable
\begin{equation} \vec{\zeta} = \sqrt{x(1-x)}\, \vec{b}_\perp = \zeta e^{i\varphi} \label{zeta} \end{equation} is introduced. This allows the LFWF to be expressed in a factorized form in terms of $x$, $\zeta$, and $\varphi$ as
\begin{equation} \Psi (x, \zeta, \varphi) = \frac{\phi (\zeta)}{\sqrt{2\pi \zeta}}\, e^{i L \varphi}\, X(x) \,, \label{full-mesonwf} 
\end{equation} 
where $\phi(\zeta)$ and $X(x) = \sqrt{x(1-x)}\, \chi(x)$ represent the transverse and longitudinal modes, respectively.

The hLFQCD, defined in the chiral limit of QCD and containing only the dynamical transverse mode, is governed by a holographic Schr\"{o}dinger-like equation~\cite{Brodsky:2006uqa,deTeramond:2005su,deTeramond:2008ht,Brodsky:2014yha}:  
\begin{equation}
    \left(-\frac{\mathrm{d}^2}{\mathrm{d} \zeta^2} + \frac{4L^2 - 1}{4 \zeta^2} + U^{\mathrm{hLF}}_\perp(\zeta)\right) \phi(\zeta) = M_\perp^2 \phi(\zeta) \;,
    \label{SEq}
\end{equation}
where the transverse confinement potential is given by  
\begin{equation}
    U_\perp^{\mathrm{hLF}}(\zeta) = \kappa^4 \zeta^2 + 2 \kappa^2 (J - 1) \;,
    \label{U-LFH}
\end{equation}
with $J = L + S$ (where $L \equiv |L_z^{\mathrm{max}}|$) denoting the total angular momentum of the meson.  
The functional form of $U_\perp^{\mathrm{hLF}}(\zeta)$ is uniquely determined~\cite{Brodsky:2013ar} by the underlying conformal symmetry
and the holographic correspondence to $\mathrm{AdS}_5$, where the light-front variable $\zeta$ is identified with the fifth dimension of AdS space. The mass scale parameter $\kappa$ defines both the confinement strength and the meson mass spectrum in the chiral limit.
The Schr\"{o}dinger equation~\eqref{SEq} can be solved analytically, yielding the mass eigenvalues
\begin{equation}
    M_{\perp}^2(n_\perp, J, L) = 4\kappa^2 \left( n_\perp + \frac{J + L}{2} \right) \;,
    \label{MTM}
\end{equation}
and the corresponding transverse wavefunctions
\begin{equation}
    \phi_{n_\perp L}(\zeta) \propto \zeta^{1/2 + L} \exp\left(-\frac{\kappa^2 \zeta^2}{2}\right) L_{n_\perp}^L(\kappa^2 \zeta^2) \;.
    \label{TMWF}
\end{equation}
A notable prediction of Eq.~\eqref{MTM} is that the lowest-lying bound state, corresponding to $n_\perp = L = S = 0$, is massless. This naturally identifies the pion as the chiral-limit Goldstone boson in QCD.

In contrast, the longitudinal mode $\chi(x)$ is not dynamical in hLFQCD and is typically set to a constant, $\chi(x) = 1$, leading to $X(x) = \sqrt{x(1-x)}$. This choice is validated by the holographic mapping of the pion’s electromagnetic and gravitational form factors between physical spacetime and $\mathrm{AdS}_5$~\cite{Brodsky:2007hb,Brodsky:2008pf}.

Substituting Eq.~\eqref{TMWF} into Eq.~\eqref{full-mesonwf} yields the chiral-limit LFWF for the pion:
\begin{equation}
    \Psi^\pi_c (x, \zeta^2) \propto \sqrt{x(1-x)} \exp\left(-\frac{\kappa^2 \zeta^2}{2}\right) \;.
    \label{pichiralwf} 
\end{equation}
Applying a two-dimensional Fourier transform to Eq.~\eqref{pichiralwf} gives the corresponding momentum-space LFWF:
\begin{align}
    \Phi^\pi_c(x, k^2_\perp ) \propto \frac{1}{\sqrt{x(1-x)}} \exp\left(-\frac{{k}_\perp^2}{2\kappa^2 x(1-x)}\right) \;,
    \label{FT}
\end{align}
where $\vec{k}_\perp$ denotes the transverse momentum of the quark, Fourier-conjugate to the transverse separation ${\vec{b}_\perp}$.

Beyond the chiral limit, Brodsky and de Téramond proposed a prescription for the longitudinal mode based on the invariant mass~\cite{Brodsky:2008pg}. While this approach has two notable limitations—it does not satisfy the Gell-Mann-Oakes-Renner (GMOR) relation and assigns the same longitudinal mode to all radially excited states—it has nonetheless been successfully applied to describe various properties of both light and heavy mesons~\cite{Gurjar:2024wpq,Brodsky:2014yha,Brodsky:2008pg,deTeramond:2021yyi,Swarnkar:2015osa,Ahmady:2020mht,Ahmady:2019hag,Ahmady:2019yvo,Ahmady:2018muv,Ahmady:2016ufq}.
An alternative approach to modeling the longitudinal dynamics uses the 't Hooft equation~\cite{tHooft:1974pnl} from large-$N_c$ QCD in (1+1) dimensions, which has been shown to reproduce the full meson mass spectrum, including the pion~\cite{Ahmady:2022dfv}, $\rho$ meson~\cite{Gurjar:2024wpq}, and both heavy-light and heavy-heavy systems~\cite{Ahmady:2021yzh}. The idea of replacing the invariant mass prescription with the 't Hooft equation was first introduced in Ref.~\cite{Chabysheva:2012fe}, which focused on predicting meson decay constants and parton distribution functions.
In parallel, recent studies~\cite{deTeramond:2021yyi,Li:2021jqb} have employed a phenomenological longitudinal confinement potential, initially proposed in Ref.~\cite{Li:2015zda} within the basis light-front quantization (BLFQ) framework. Although these works primarily address the chiral limit and chiral symmetry breaking, Ref.~\cite{deTeramond:2021yyi} also explores applications to ground-state heavy mesons and connects their approach to the 't Hooft equation.
These developments reflect a growing interest in incorporating longitudinal dynamics into hLFQCD~\cite{Li:2021jqb,deTeramond:2021yyi,Lyubovitskij:2022rod,Weller:2021wog,Ahmady:2021lsh,Rinaldi:2022dyh}.


Starting from the QCD Lagrangian in $(1+1)$ dimensions and applying the large-$N_c$ approximation, 't Hooft derived a Schr\"odinger-like equation for the meson wavefunction~\cite{tHooft:1974pnl}:  
\begin{align}
 	\left(\frac{m_1^2}{x}+\frac{m_2^2}{1-x}\right)\chi(x) + U_\parallel (x)\chi(x) = M^2_\parallel \chi(x) \;,
 	\label{tHooft}
\end{align}
where \(m_1\) and \(m_2\) denote the masses of the constituents, here the gluon and the spectator, respectively, and the longitudinal potential is given by
\begin{equation}
	U_\parallel(x)\chi(x) = \frac{g^2}{\pi} \, \mathcal{P} \int {\rm d}y \, \frac{\chi(x)-\chi(y)}{(x-y)^2} \;,
	\label{tHooft-potential}
\end{equation}
with $g$ denoting the longitudinal confinement scale and $\mathcal{P}$ representing the Cauchy principal value.  
Unlike the holographic light-front Schr\"odinger equation, the 't Hooft equation does not admit an analytical solution and must be solved numerically. Here, we follow the matrix method outlined in Ref.~\cite{Chabysheva:2012fe}.


Combining the holographic Schr\"odinger equation for the transverse mode with the 't Hooft equation for the longitudinal mode, the meson mass spectrum is given by:
\begin{align}
	M^2(n_\perp, n_\parallel, J, L) = 4\kappa^2\left(n_\perp + \frac{J+L}{2}\right) + M^2_\parallel(n_\parallel, m_1, m_2, g) \;,
	\label{totalmass}
\end{align}
where $n_{\parallel}$ is the longitudinal quantum number. For the ground state ($n_{\perp} = L = S = n_{\parallel} = 0$), this formula reproduces the physical pion mass, in contrast to Eq.~\eqref{MTM}, where the pion is massless in the chiral limit. This shows that the 't Hooft equation accounts for the full mass of the pion.

\begin{table}
\caption{Quantum numbers and masses of the pion family. The transverse mass \( M_\perp \) and longitudinal mass \( M_\parallel \) are obtained by solving the holographic Schr\"odinger equation and the 't Hooft equation, respectively. The total mass \( M \) of the system is then computed using Eq.~\eqref{totalmass}.}
\centering
\begin{tabular}{|c c c c c c c c|}
\hline
$J^{P(C)}$ & Name & $n_\perp$ & $n_\parallel$ & $L$ & $M_\perp$ (MeV) & $M_\parallel$ (MeV) & $M$ (MeV) \\
\hline
$0^{-}$ & $\pi(140)$ & 0 & 0 & 0 & 0 & 135 & 135 \\
$0^{-+}$ & $\pi(135)$ & 0 & 0 & 0 & 0 & 135 & 135 \\
$1^{+-}$ & $b_{1}(1235)$ & 0 & 2 & 1 & 1046 & 303 & 1089 \\
$0^{-+}$ & $\pi(1300)$ & 1 & 2 & 0 & 1046 & 303 & 1089 \\
$2^{-+}$ & $\pi_{2}(1670)$ & 0 & 4 & 2 & 1479 & 409 & 1534 \\
$0^{-+}$ & $\pi(1800)$ & 2 & 4 & 0 & 1479 & 409 & 1534 \\
$2^{-+}$ & $\pi_2(1880)$ & 1 & 6 & 2 & 1811 & 492 & 1877 \\
\hline
\end{tabular}
\label{Spectroscopy}
\end{table}

\section{Pion Mass Spectroscopy and Light-Front Wavefunctions}

We employ a physical gluon mass, setting $m_1 = 0$. To fit the spectroscopic data of the pion family, we fix the three free parameters of our model: the longitudinal confinement scale $g$, the transverse confinement scale $\kappa$, and the spectator mass $m_2$. We take $m_2 = 0.092$ GeV, which is twice the light quark mass used in  hLFQCD with the Brodsky--de Téramond ansatz~\cite{Brodsky:2014yha}. For the transverse confinement scale, we use the universal value $\kappa = 0.523$ GeV, consistent across the full hadron spectrum~\cite{Ahmady:2021yzh}. In contrast, the longitudinal confinement scale $g$ is not universal; here, we use $g = 0.109$ GeV~\cite{Ahmady:2021yzh}. Following Refs.~\cite{Ahmady:2021lsh,Ahmady:2021yzh,Ahmady:2022dfv}, the parity and charge conjugation quantum numbers of mesons are given by
\begin{equation}
    P = (-1)^{L+1}, \quad C = (-1)^{L + S + n_\parallel}.
    \label{pcrules}
\end{equation}

\begin{figure}[hbt!]
	\begin{center}
		\includegraphics[width=0.6\linewidth]{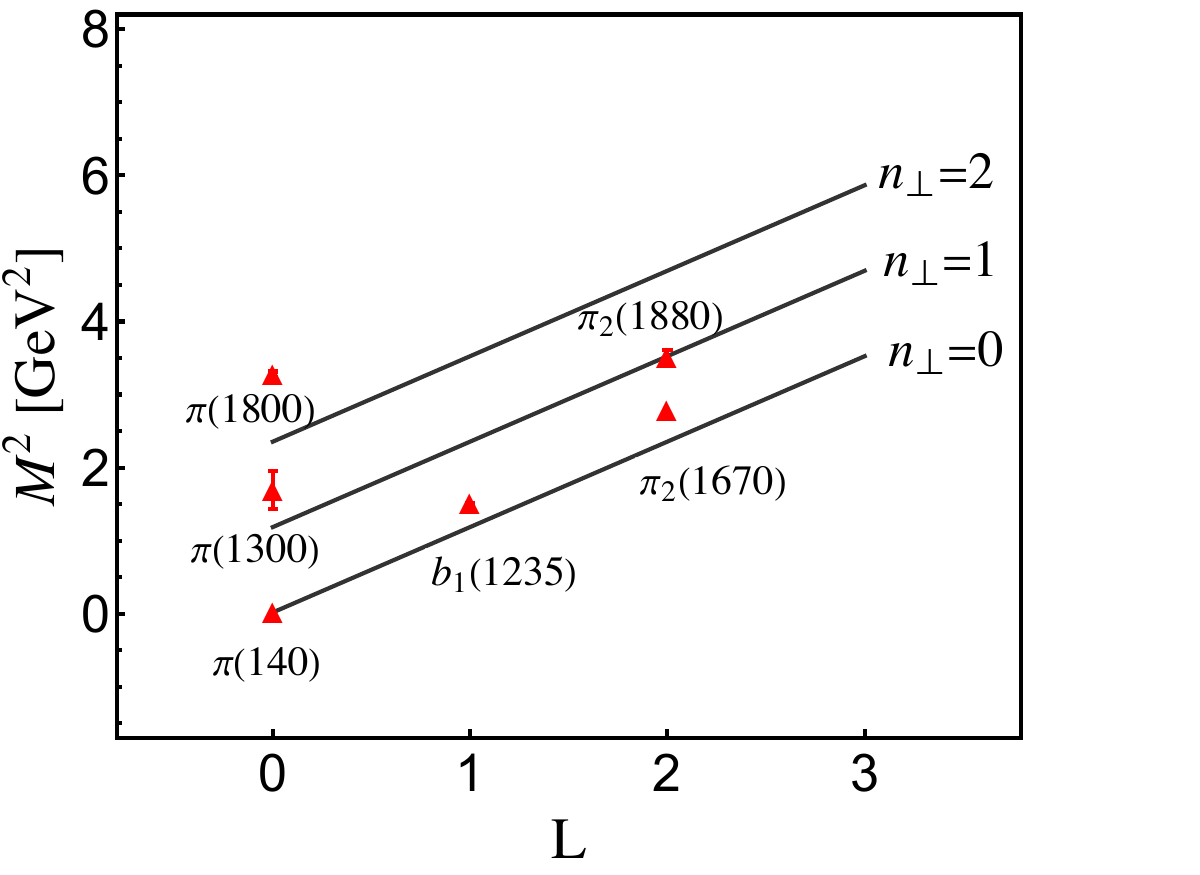}
		\caption{Our Regge trajectories for the pion family using the physical gluon mass $m_{1}=0$, the spectator mass $m_{2}=0.092$ GeV,  $g=0.109$ GeV and $\kappa=0.523$ GeV.}
		\label{pionmass}
	\end{center}
\end{figure}

We present the computed masses of the pion and its excitations in Table~\ref{Spectroscopy}. Our results (last column) show good agreement with experimental values (second column, in parentheses). Notably, the condition $n_\parallel \ge n_\perp + L$~\cite{Ahmady:2021yzh}, observed in Table~\ref{Spectroscopy}, appears to hold throughout the hadron spectrum. The corresponding Regge trajectories for the pion family are shown in Fig.~\ref{pionmass}.


\begin{figure}[hbt!]
	\centering
\includegraphics[width=0.5\textwidth]{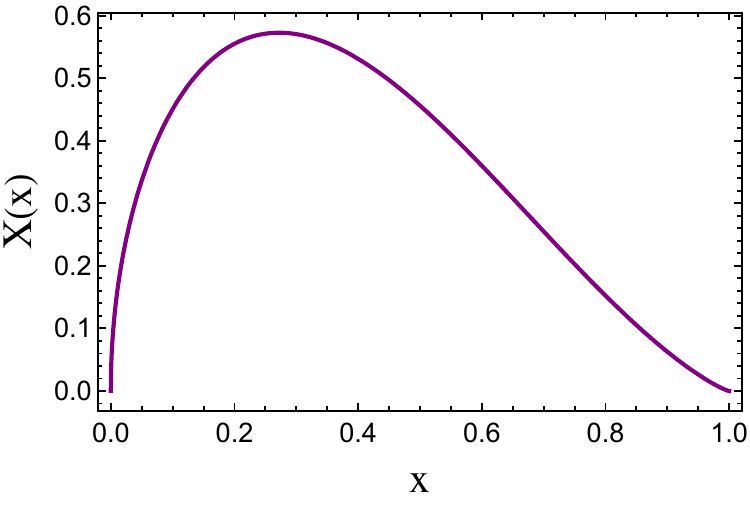}
\caption{The longitudinal wave function $X(x)$ of gluon in the pion.}\label{LWF}
\end{figure}

Although the longitudinal mode $\chi(x)$ cannot be determined analytically, it can be well approximated by:
\begin{eqnarray}
\chi(x) \approx a_0 + a_1 x+a_2 x^2+a_3 x^3 +a_4 x^4 +a_5 x^5 \;,
\label{eq:tHooftanaly}
\end{eqnarray}
where $a_i$ are constituent-mass-dependent parameters that vanish in the chiral limit except $a_0 (=1)$. With the chosen parameters, we find that for the pion ground state, $\lbrace a_0, a_1,a_2,a_3,a_4,a_5 \rbrace \approx \lbrace1.57, -0.49, -2.47, 1.53, 0.50, -0.62 \rbrace$.  Figure~\ref{LWF} illustrates the dynamical longitudinal wavefunction of the ground-state pion: $X(x)=\sqrt{x(1-x)}\chi(x)$, where $x$ is momentum fraction carried by the gluon.


Incorporating both the transverse mode from the light-front Schr\"odinger equation and the longitudinal mode from the 't Hooft equation, the spin-independent LFWF of the meson takes the form:
\begin{eqnarray}
\Phi^\pi(x, {k}^2_\perp) = \mathcal{N} \frac{\chi(x)}{\sqrt{x(1-x)}} \exp\left(-\frac{{k}_\perp^2}{2\kappa^2 x(1-x)}\right) \;,
\label{eq:totalhwf}
\end{eqnarray}
where $\chi(x)$ is the longitudinal mode obtained from the 't Hooft equation and $\mathcal{N}$ is a normalization constant.

In our gluon-spectator model, we assume that the spectator has spin 1, same as the gluon.
The two-particle Fock-state expansion for the pion's orbital angular momentum component, $L^z = 0$, in a frame where the pion's transverse momentum vanishes, i.e., $P \equiv \big(P^+, \vec{0}_\perp, \frac{M^2}{P^+}\big)$, is expressed as~\cite{Pasquini:2023aaf}
	\begin{align}\label{state}
		|\pi(P)\rangle
		= &\int \frac{\mathrm{d} x\,\mathrm{d}^2 \vec{k}_\perp }{16 \pi^3 \sqrt{x(1-x)}} \Big[\psi_{-1,+1}^\pi\left(x, \vec{k}_\perp\right)\left|-1,+1 ; x P^{+}, \vec{k}_\perp\right\rangle+\psi_{+1,-1}^\pi\left(x, \vec{k}_\perp\right)\left|+1,-1 ; x P^{+}, \vec{k}_\perp\right\rangle\Big].
	\end{align}	
For nonzero transverse momentum of the pion, i.e., $\vec{P}_\perp \neq 0$, the physical transverse momenta of the gluon and spectator are given by $\vec{p}_\perp^{\,g} = x\vec{P}_\perp + \vec{k}_\perp$ and $\vec{p}_\perp^{\,s} = (1-x)\vec{P}_\perp - \vec{k}_\perp$, respectively, where $\vec{k}_\perp$ represents the relative transverse momentum of the gluon. The LFWFs, denoted by $\psi_{\lambda_g,\lambda_s}^{\pi}(x,\vec{k}_\perp)=\Phi^\pi(x, \vec{k}_\perp)\lambda_g\delta_{\lambda_g,-\lambda_s}$ correspond to the two-particle state $|\lambda_{g}, \lambda_s; xP^{+}, \vec{k}_\perp \rangle$, with the gluon's helicity $\lambda_g = \pm 1$, and the spectator's helicity $\lambda_s = \pm 1$. 
The LFWF is normalized as follows:
\begin{eqnarray}
\sum_{\lambda_g,\lambda_s}\int \frac{{\mathrm d} x\,{\mathrm d}^2 \vec{k}_{\perp}}{16\pi^3} \left|\psi^\pi_{\lambda_g, \lambda_s}(x, \vec{k}_\perp)\right|^{2} = 1 \;.
\label{eq:normalization}
\end{eqnarray}

\section{Gluon generalized parton distributions}
The pion gluon GPDs are defined through the off-forward matrix elements of the bilocal gluon tensor operator between pion states~\cite{Diehl:2003ny}:
\begin{align}
H^{g}_\pi(x,\,\xi,\,t) = \frac{1}{\bar{P}^+}\int \frac{{\rm d}z^-}{2\pi} e^{ix\bar{P}^+z^-}  \langle \pi (P^{\prime}) |\, F^{+i} \left(-\frac{z}{2} \right) F^{+i}  \left(\frac{z}{2} \right) \,| \pi (P) \rangle \Big{|}_{\substack{z^+=0\\z^{\perp}=0}},
\end{align}
where $F^{+\mu}(z) = \partial^{+}A^{\mu}(z)$ corresponds to the gluon field tensor in the light-cone gauge $A^+=0$. In the symmetric frame, the average momentum of pion $\bar{P}= \frac{1}{2} (P^{\prime}+P)$, while momentum transfer $\Delta=(P^{\prime}-P)$. The initial and final four momenta of the pion are then given by
\begin{equation}
\begin{aligned}
P &\equiv \left((1+\xi)\bar{P}^+,\frac{M^2+\Delta_\perp^2/4}{(1+\xi)P^+},-\vec{\Delta}_\perp/2\right),\\
P^{\prime} &\equiv \left((1-\xi)\bar{P}^+,\frac{M^2+\Delta_\perp^2/4}{(1-\xi)P^+},\vec{\Delta}_\perp/2\right), \label{Ppp}
\end{aligned}
\end{equation}
where $\xi=- \Delta^+/2\bar{P}^+$ and $t= \Delta^2$. One can derive the following relation explicitly from $\Delta^-$,
\begin{align}
- t= \frac{4 \xi^2 M^2 + \Delta_\perp^2}{(1-\xi^2)}\,. \label{mt_def}
\end{align}

The overlap representation of the GPD in terms of the LFWFs is expressed as
\begin{eqnarray}
H^{g}_\pi(x,\,\xi,\,t) =\sum_{\lambda_g,\lambda_s}\int \left[{\rm d}\mathcal{X} \,{\rm d}\mathcal{K}_\perp\right]\, \psi_{\lambda_g,\lambda_s}^{\pi*}(x^{\prime\prime}, \vec{k}_\perp^{\prime\prime}) \psi_{\lambda_g,\lambda_s}^{\pi}(x^{\prime}, \vec{k}_\perp^{\prime}),
\label{eq:GPD_WF}
\end{eqnarray}
where 
\begin{align}
\left[{\rm d}\mathcal{X} \,{\rm d}\mathcal{K}_\perp\right]\equiv&\sqrt{x^2-\xi^2}(\sqrt{1-\xi^2}\ )^{-1}\prod_{i=1}^2 \left[\frac{{\rm d}x_i{\rm d}^2 \vec{k}_{i\perp}} {16\pi^3}\right]\delta(x-x_1) 16 \pi^3 \delta \left(1-\sum_{i=1}^{2} x_i\right) \delta^2 \left(\sum_{i=1}^{2}\vec{k}_{i\perp}\right), \nonumber  
\end{align} 
and the light-front momenta are $x^{\prime}_1=\frac{x_1+\xi}{1+\xi}$; $\vec{k}^{\prime}_{1\perp}=\vec{k}_{1\perp}+(1-x^{\prime})\frac{\vec{\Delta}_{\perp}}{2}$ for the initial gluon ($i=1$) and $x^{\prime}_i=\frac{x_i}{1+\xi}; ~\vec{k}^{\prime}_{i\perp}=\vec{k}_{i\perp}-{x_i^{\prime}} \frac{\vec{\Delta}_{\perp}}{2}$ for the initial spectator  ($i\ne1$), and
$x^{\prime\prime}_1=\frac{x_1-\xi}{1-\xi}$; $\vec{k}^{\prime\prime}_{1\perp}=\vec{k}_{1\perp}-(1-x^{\prime\prime})\frac{\vec{\Delta}_{\perp}}{2}$ for the final gluon and $x^{\prime\prime}_i=\frac{x_i}{1-\xi}; ~\vec{k}^{\prime \prime}_{i\perp}=\vec{k}_{i\perp}+{x_i^{\prime\prime}} \frac{\vec{\Delta}_{\perp}}{2}$ for the final spectator.
Note that, in this work, we limit ourselves to the Dokshitzer-Gribov-Lipatov-Altarelli-Parisi (DGLAP) region, $\xi<x<1$, where the number of partons in the initial and the final states remains conserved. 


\begin{figure}[hbt!]
     \centering
     \begin{subfigure}[b]{0.47\textwidth}
         \centering
         \includegraphics[width=\textwidth]{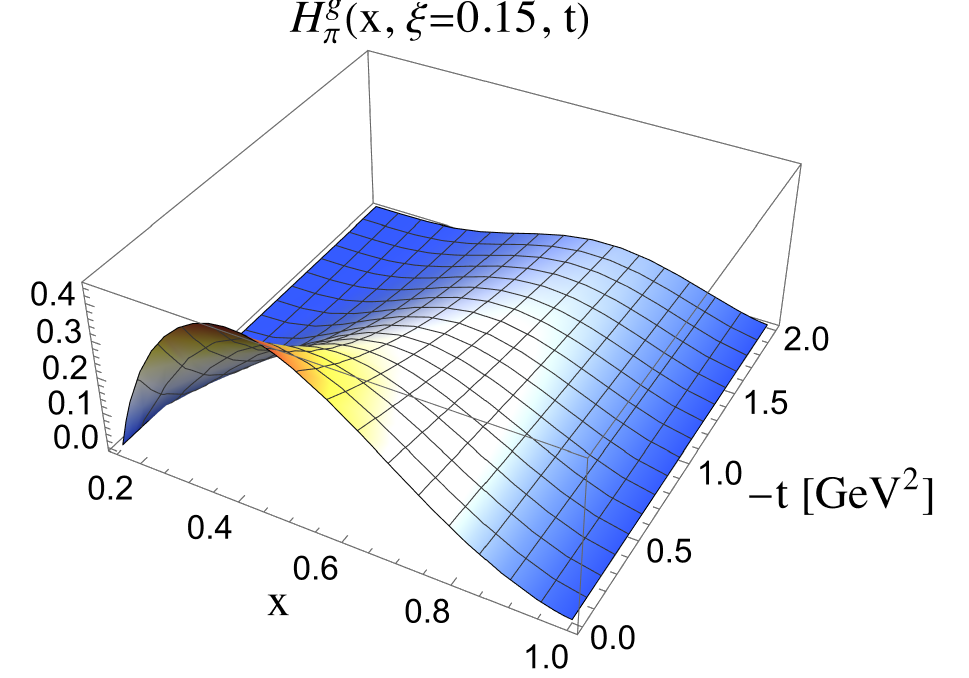}
         \caption{}
         \label{fig:gpd-fixed-zeta}
     \end{subfigure}
     \begin{subfigure}[b]{0.38\textwidth}
         \centering
         \includegraphics[width=\textwidth]{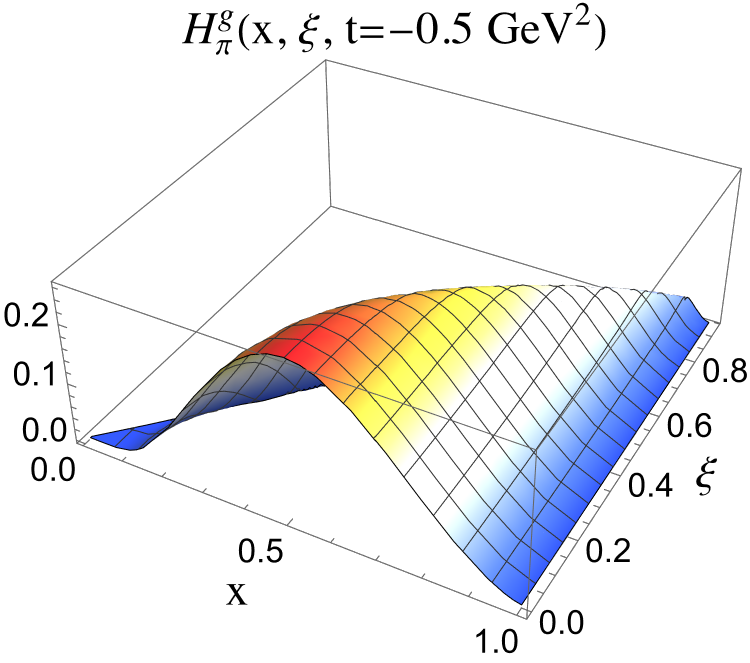}
         \caption{}
         \label{fig:gpd-fixed-t}
     \end{subfigure}
             \caption{
             The gluon GPD of the pion in the DGLAP region ($\xi < x < 1$) as a function of (a) $
x$ and $-t$ [in GeV$^2$] at fixed skewness, $\xi=0.15$, and (b) $
x$ and $\xi$ at fixed $-t=0.5$ GeV$^2$.}
        \label{gpd:zero}
\end{figure}

We present the pion's gluon GPD as a function of $x$ and $-t$ at fixed $\xi = 0.15$ in the left panel of Fig.~\ref{gpd:zero}. Physically, this corresponds to probing the pion off-forward, so the distribution is shifted toward configurations where the gluon carries slightly different longitudinal momentum after the interaction, leading to a skewed distribution. The distribution peaks when there is no transverse momentum transfer and the gluon carries less than 50\% of the pion’s longitudinal momentum. As $-t$ increases, the peak shifts to higher $x$ values, and the overall magnitude decreases. At large $x$, the GPDs decay and become nearly independent of $-t$. This behavior is consistent with trends observed in gluon GPDs~\cite{Lin:2023ezw,Zhang:2025nll,Lin:2024ijo,Chakrabarti:2024hwx} and quark GPDs in both the proton~\cite{Liu:2024umn,Ji:1997gm,Boffi:2002yy,Boffi:2003yj,Mondal:2015uha,Mondal:2017wbf,Freese:2020mcx} and the pion~\cite{Kaur:2018ewq,Kaur:2020vkq}.

The right panel of Fig.~\ref{gpd:zero} shows the pion’s gluon GPD as a function of $x$ and $\xi$ at fixed $t = -0.5$ GeV$^2$. The distribution peaks at low $x$ ($< 0.5$) and shifts toward higher $x$ with increasing $\xi$, while the peak height diminishes. At large $x$, the distributions again become insensitive to $\xi$. These features are model- and parton-independent and consistent with parton distributions across various QCD-inspired models~\cite{Ji:1997gm,Boffi:2002yy,Boffi:2003yj,Mondal:2015uha,Mondal:2017wbf,Freese:2020mcx,Pasquini:2005dk,Chakrabarti:2015ama}.

We emphasize that the current study is limited to the DGLAP region of the gluon GPD, and the ERBL domain has not been reconstructed. Therefore, the polynomiality property of Mellin moments is not exactly fulfilled within the present framework, and any comparisons relying on full $x$-support moments extend beyond our current scope.

To illustrate the impact of the missing ERBL contribution, we perform a minimal extension by embedding our DGLAP-region GPD into a simple double-distribution (DD) model supplemented with a phenomenological D-term. This construction approximately restores the polynomial behavior of the Mellin moments, as shown in Fig.~\ref{polynomiality} in the Appendix~\ref{appendix} confirming the internal consistency of our DGLAP input and providing a useful baseline for future work incorporating a complete ERBL description.


\subsection{Gluon PDF}
\begin{figure}[hbt!]
     \centering
     \begin{subfigure}[b]{0.47\textwidth}
         \centering
         \includegraphics[width=\textwidth]{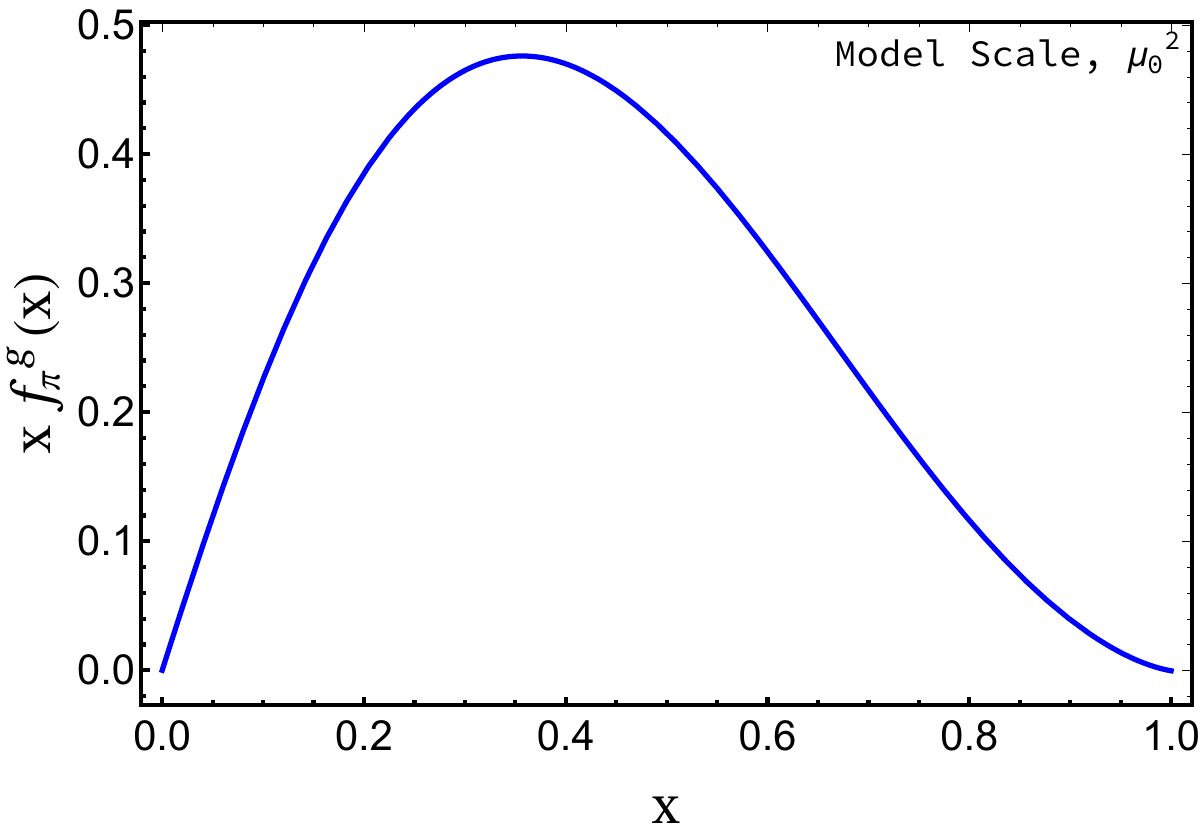}
         \caption{}
         \label{fig:pdf-initial}
     \end{subfigure}
     \begin{subfigure}[b]{0.47\textwidth}
         \centering
         \includegraphics[width=\textwidth]{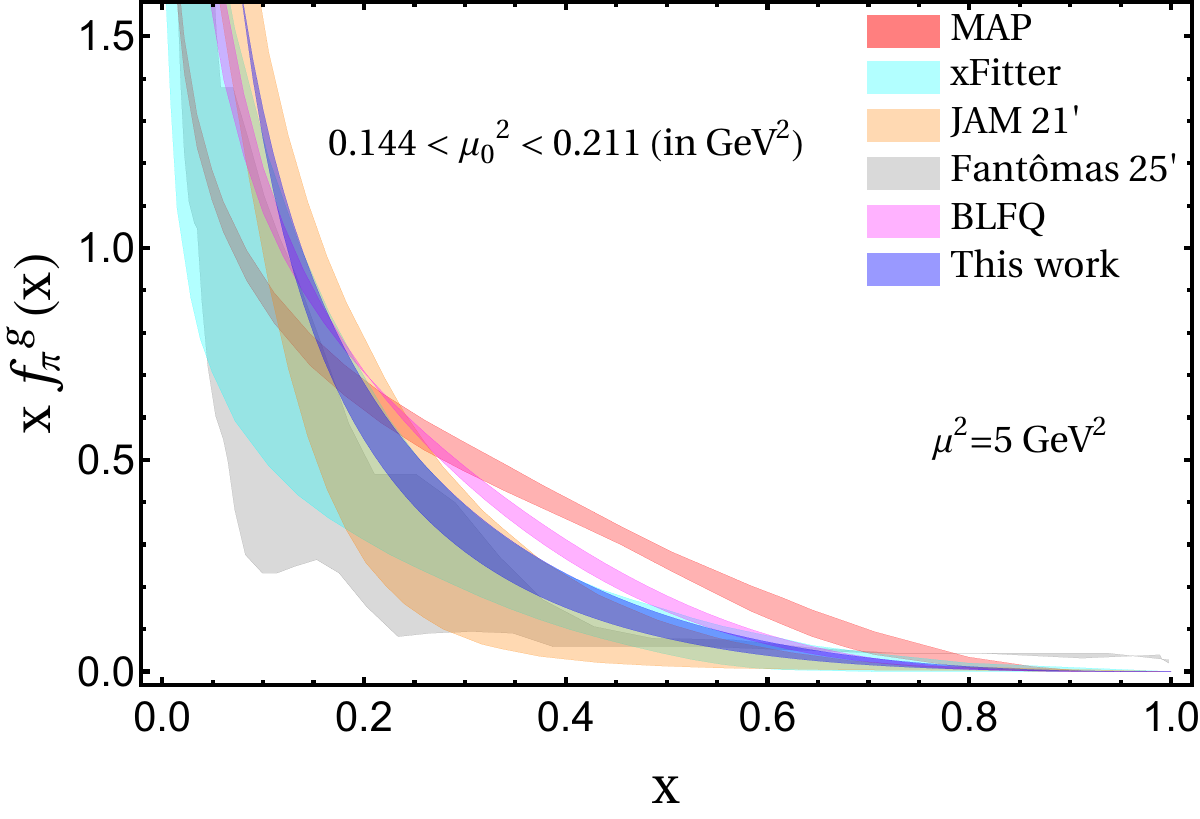}
         \caption{}
         \label{fig:pdf-final}
     \end{subfigure}
            \caption{
The pion’s gluon PDF at (a) the model scale (\(\mu_0^2 = 0.176\) GeV$^2$) and (b) the scale \(\mu^2 = 5~\mathrm{GeV}^2\) evolved via DGLAP, varying the model scale within \(0.144 < \mu_0^2 < 0.211~\mathrm{GeV}^2\). The evolved results are compared with global QCD fits: JAM~\cite{Barry:2021osv} (orange band), xFitter~\cite{Novikov:2020snp} (cyan band), Fant$\hat{\rm o}$mas~\cite{Kotz:2025lio} (gray band), a phenomenological model by the MAP Collaboration~\cite{Pasquini:2023aaf} (red band), and a Hamiltonian-based BLFQ framework~\cite{Lan:2024ais} (magenta band).}
        \label{fig:PDF}
\end{figure}

In the forward limit, the gluon GPD reduces to the unpolarized gluon PDF, $H^{g}_{\pi}(x,\,0,\,0) = x f^g_{\pi}(x)$~\cite{Diehl:2003ny}. The gluon PDF at the model scale is shown in the left panel of Fig.~\ref{fig:PDF}. We also perform QCD evolution using the next-to-next-to-leading order (NNLO) DGLAP equations to obtain the PDF at the scale \( \mu^2 = 5\,\text{GeV}^2 \), enabling comparison with global fits~\cite{Barry:2021osv,Novikov:2020snp,Kotz:2025lio}, a phenomenological model~\cite{Pasquini:2023aaf}, and theoretical predictions from the Hamiltonian-based Basis Light-Front Quantization (BLFQ) approach~\cite{Lan:2024ais}.
 We determine the model scale $\mu_0$ by requiring that the evolved gluon distribution reproduces the first moment obtained from lattice QCD simulations, $\langle x \rangle^{g}_\pi = 0.61 \pm 0.09$ at $\mu^2 = 4~\text{GeV}^2$~\cite{Shanahan:2018pib}. This constraint yields a model scale of $\mu_0=0.42 \pm 0.04~\text{GeV}$. In solving the DGLAP equations, we impose the condition that the running coupling $\alpha_s(\mu^2)$ saturates in the infrared, with a maximum cutoff value of $\max\{\alpha_s\} \sim 1$~\cite{Lan:2020fno,Lan:2019vui}.

 \begin{figure}[hbt!]
     \centering
     \begin{subfigure}[b]{0.47\textwidth}
         \centering
         \includegraphics[width=\textwidth]{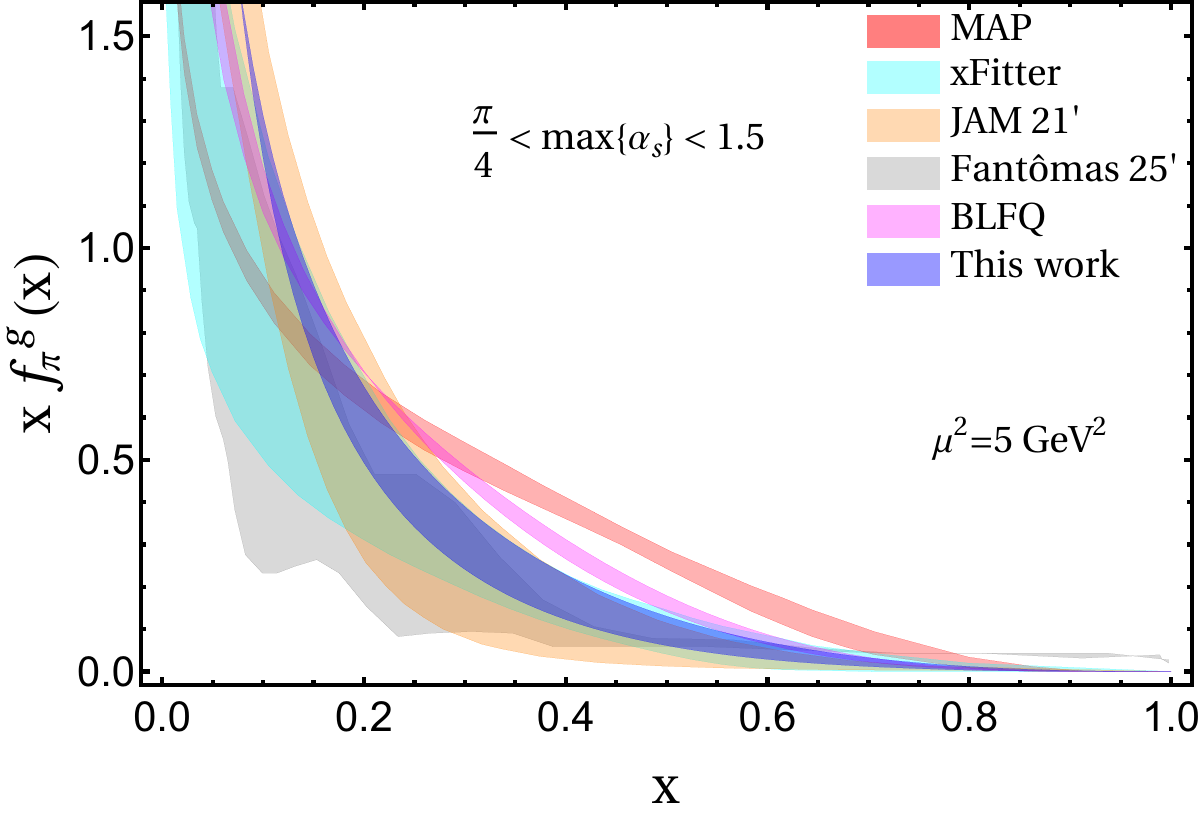}
         \caption{}
         \label{fig:pdf-initial_as}
     \end{subfigure}
     \begin{subfigure}[b]{0.47\textwidth}
         \centering
         \includegraphics[width=\textwidth]{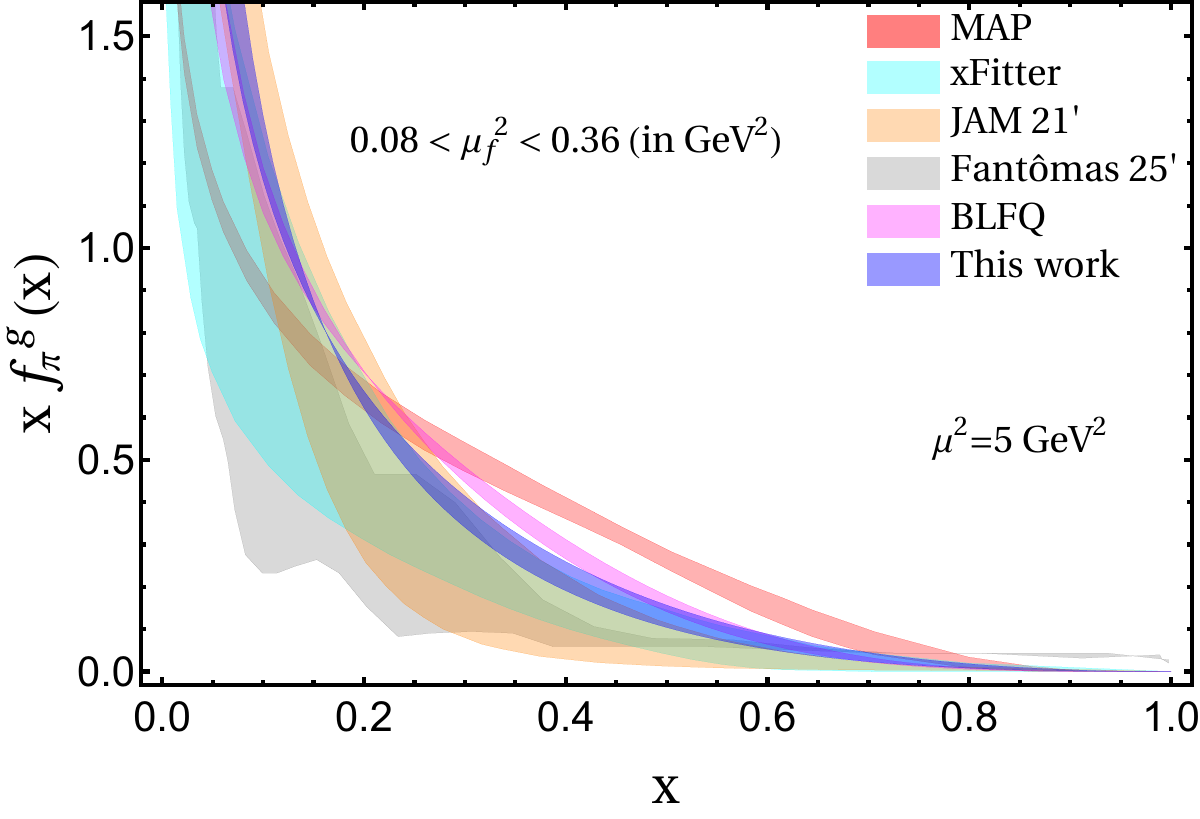}
         \caption{}
         \label{fig:pdf-final_muf}
     \end{subfigure}
             \caption{
Gluon PDF of the pion at the scale $5~\mathrm{GeV}^2$ evolved via DGLAP with different IR-freezing prescriptions: 
(a) varying $\alpha_s^{\mathrm{max}}$ in the range $\frac{\pi}{4} < \alpha_s^{\mathrm{max}} < 1.5$; 
(b) varying the freeze scale $\mu_f^2$ in $0.08 < \mu_f^2 < 0.36~\mathrm{GeV}^2$. 
Results are compared with global QCD fits JAM~\cite{Barry:2021osv} (orange), 
xFitter~\cite{Novikov:2020snp} (cyan), and Fant$\hat{\mathrm{o}}$mas~\cite{Kotz:2025lio} (gray), 
as well as the MAP model~\cite{Pasquini:2023aaf} (red) and BLFQ framework~\cite{Lan:2024ais} (magenta).
}
        \label{fig:PDFas}
\end{figure}
The right panel of Fig.~\ref{fig:PDF} shows our results for the pion's gluon PDF after QCD evolution. The error bands reflect a 10\% uncertainty in the initial scale. As shown, our results show similar behavior to existing global fits~\cite{Barry:2021osv,Novikov:2020snp,Kotz:2025lio} and phenomenological predictions~\cite{Pasquini:2023aaf,Lan:2024ais}, with better agreement observed with the JAM21 global QCD analysis~\cite{Barry:2021osv}.

Figure~\ref{fig:PDFas} shows the pion’s gluon PDF after QCD evolution to the scale $\mu^2 = 5~\mathrm{GeV}^2$, with the error bands representing uncertainties from different IR-freezing prescriptions. In Fig.~\ref{fig:pdf-initial_as}, we display the results obtained by varying the maximum value of the coupling, $\alpha_s^{\mathrm{max}}$, within the range $\frac{\pi}{4} < \alpha_s^{\mathrm{max}} < 1.5$, while evolving the pion’s gluon distribution from the model scale $\mu_0^2 = 0.176~\mathrm{GeV}^2$ to $\mu^2 = 5~\mathrm{GeV}^2$. Figure~\ref{fig:pdf-final_muf} illustrates the corresponding effects when varying the freeze scale in the range $0.08 < \mu_f^2 < 0.36~\mathrm{GeV}^2$. The resulting variations remain moderate and lie within the overall model uncertainty.

 \begin{figure}[hbt!]
     \centering
     \begin{subfigure}[b]{0.47\textwidth}
         \centering
         \includegraphics[width=\textwidth]{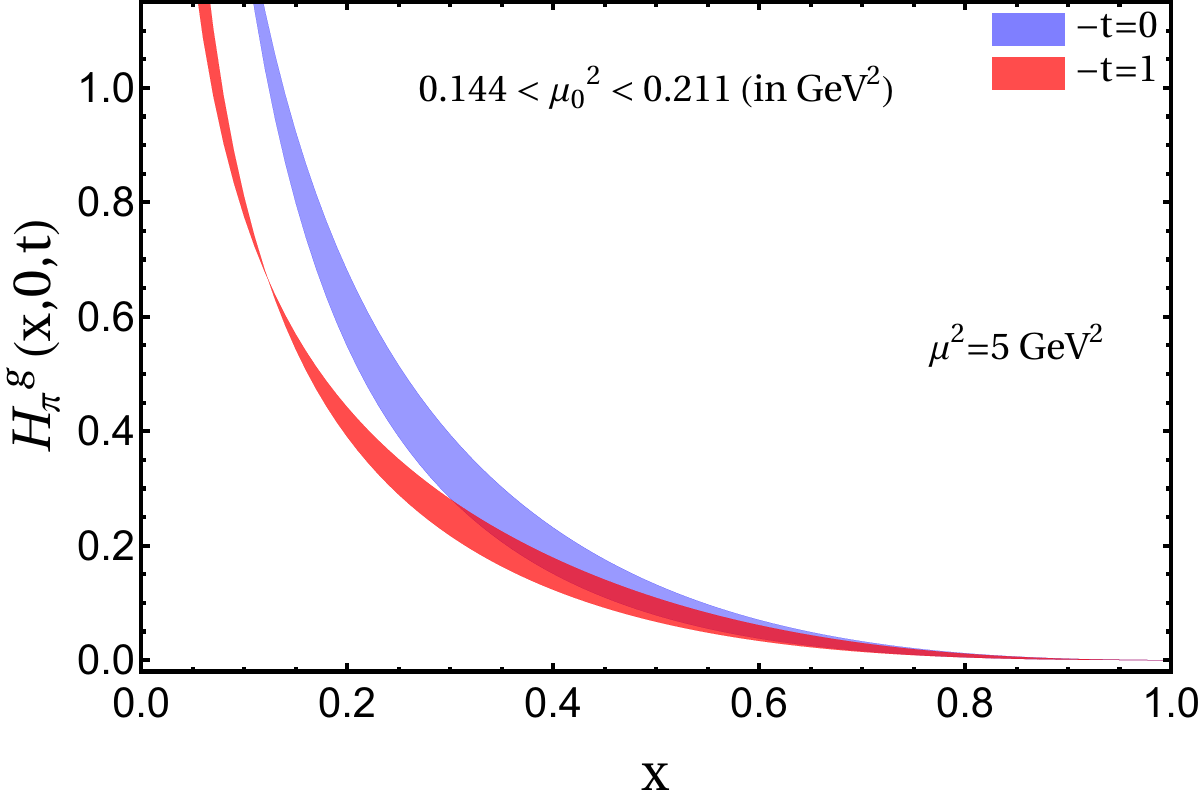}
         \caption{}
         \label{fig:GPD-evolve_scale}
     \end{subfigure}
     \begin{subfigure}[b]{0.47\textwidth}
         \centering
         \includegraphics[width=\textwidth]{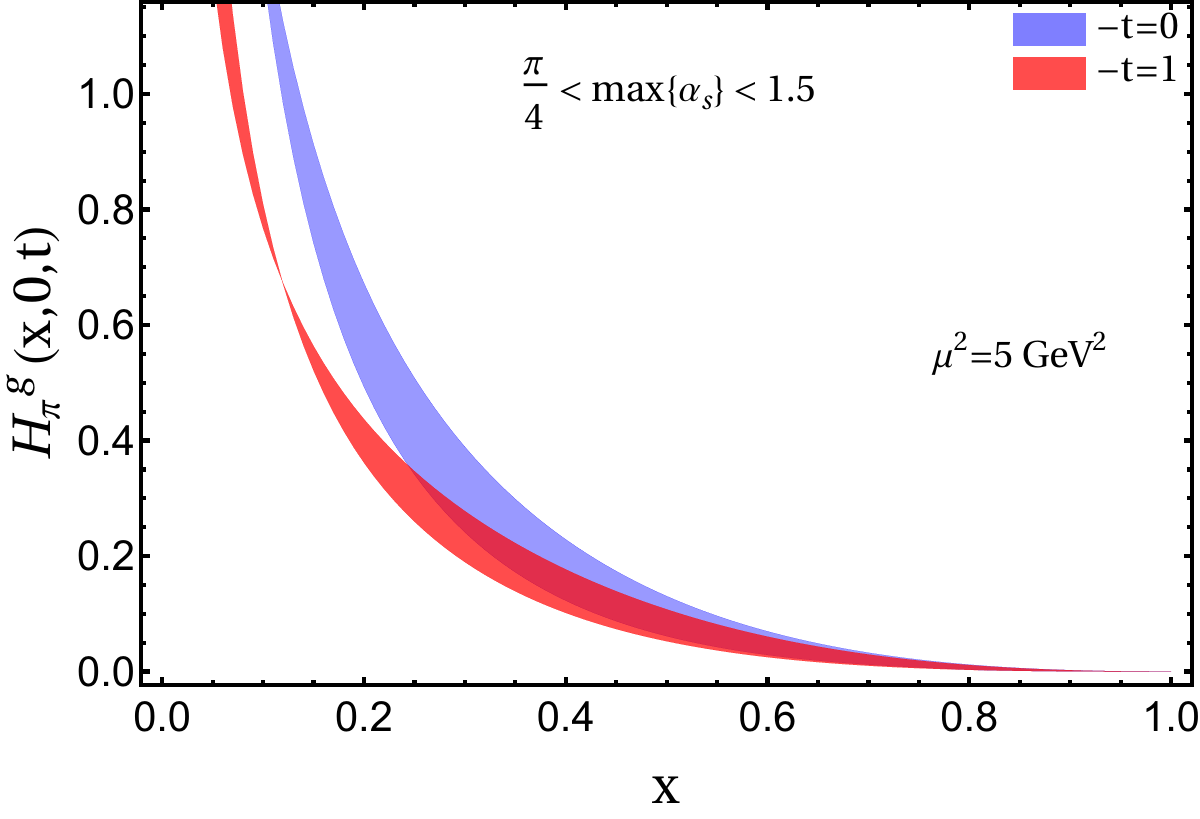}
         \caption{}
         \label{fig:GPD-evolve_as}
     \end{subfigure}
     \begin{subfigure}[b]{0.47\textwidth}
         \centering
         \includegraphics[width=\textwidth]{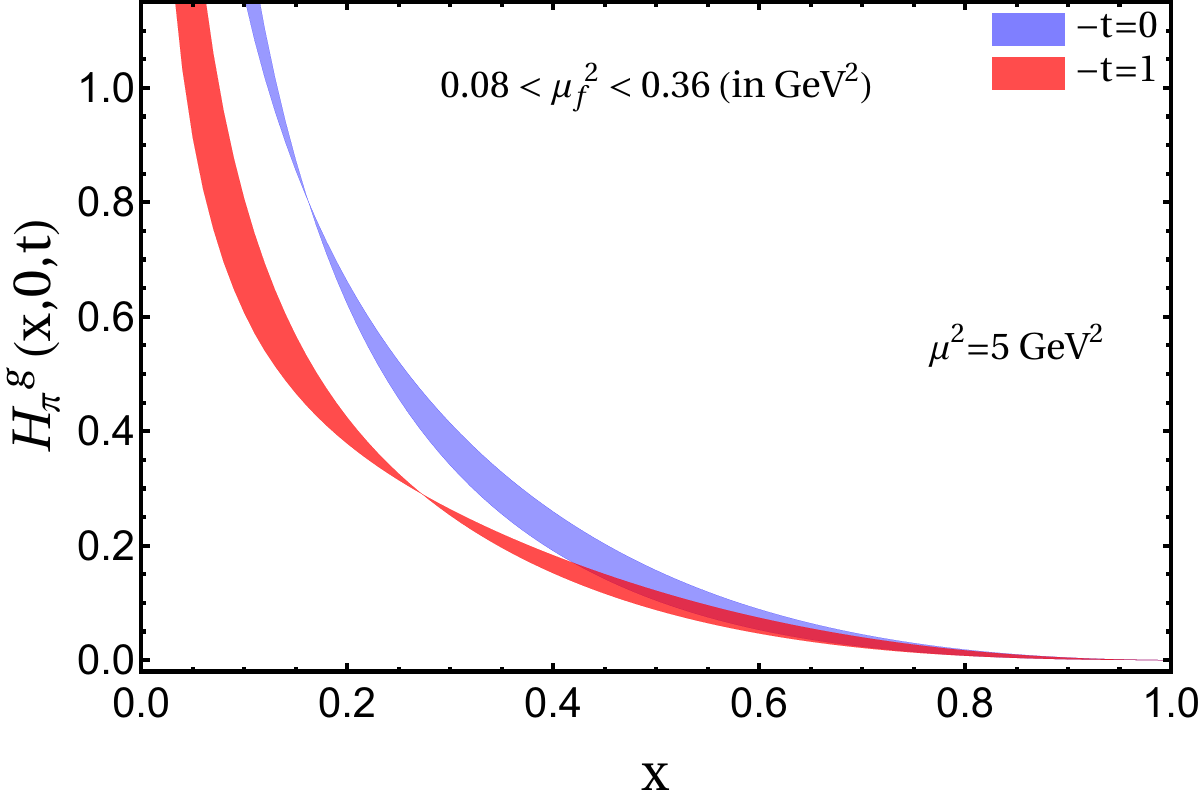}
         \caption{}
         \label{fig:GPD-evolve_muf}
     \end{subfigure}
             \caption{
Gluon GPD of the pion at the scale $5~\mathrm{GeV}^2$ evolved via DGLAP with different model scale and  IR-freezing prescriptions: (a) varying the model scale $\mu_0^2$ in $0.144 < \mu_0^2 < 0.211~\mathrm{GeV}^2$;
(b) varying $\alpha_s^{\mathrm{max}}$ in the range $\frac{\pi}{4} < \alpha_s^{\mathrm{max}} < 1.5$; 
(c) varying the freeze scale $\mu_f^2$ in $0.08 < \mu_f^2 < 0.36~\mathrm{GeV}^2$. The blue and red bands correspond to the results for $-t=0$ and $-t=1$ GeV$^2$, respectively.
}

        \label{fig:GPDas}
\end{figure}

Furthermore, we perform the QCD evolution of the pion’s gluon GPD, $H_\pi^g(x, 0, t)$, to a higher scale $\mu^2$ following the same prescriptions used for the gluon PDF. The evolved GPD results are shown in Fig.~\ref{fig:GPDas} for a momentum transfer of $-t = 1~\mathrm{GeV}^2$ and compared with the corresponding evolved PDF case, i.e., $-t = 0$.

As the gluon is probed at shorter distances and the momentum transfer $-t$ increases, the distribution shifts toward lower values of $x$ and decreases in magnitude. We also observe a pronounced enhancement at small $x$, indicating that gluons increasingly dominate the low-$x$ region as the scale evolves.

\subsection{Gluon distributions in transverse position space}
We compute the gluon distribution in transverse impact-parameter space by taking the two-dimensional Fourier transform of the momentum-space GPD with respect to the transverse momentum transfer~\cite{Diehl:2002he,Burkardt:2002hr},
\begin{equation}
H^g_\pi(x,\xi,{b}_\perp)\;=\;\frac{1}{\left(1-\xi^2\right)}\int \frac{{\rm d}^2 \vec{\Delta}_\perp}{(2\pi)^2}\,e^{-i\frac{\vec{\Delta}_\perp}{\left(1-\xi^2\right)}\cdot\vec{b}_\perp}\,
H^g_\pi \!\left(x,\xi,t=-\frac{\Delta_\perp^2+4\xi^2M^2}{1-\xi^2}\right),
\end{equation}
where the mapping $(x,\vec{\Delta}_\perp/(1-\xi^2)) \to (x,\vec{b}_\perp)$ accounts for the Lorentz-invariant skewness parameter $\xi$ and its kinematic relation to the momentum transfer $\vec{\Delta}_\perp$~\cite{Diehl:2003ny,Belitsky:2005qn}. Here, $b_\perp=|\vec{b}_\perp|$ denotes the transverse distance of the gluon relative to the center of momentum of the pion and provides a spatial image of the gluon distribution.

\begin{figure}[hbt!]
     \centering
     \begin{subfigure}[b]{0.47\textwidth}
         \centering
         \includegraphics[width=\textwidth]{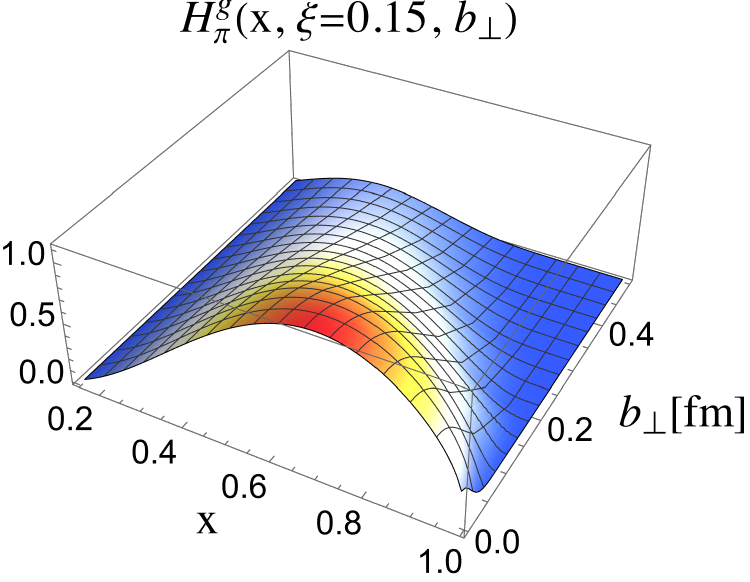}
         \caption{}
         \label{fig:gpd-position-fixed-zeta}
     \end{subfigure}
     \begin{subfigure}[b]{0.45\textwidth}
         \centering
         \includegraphics[width=\textwidth]{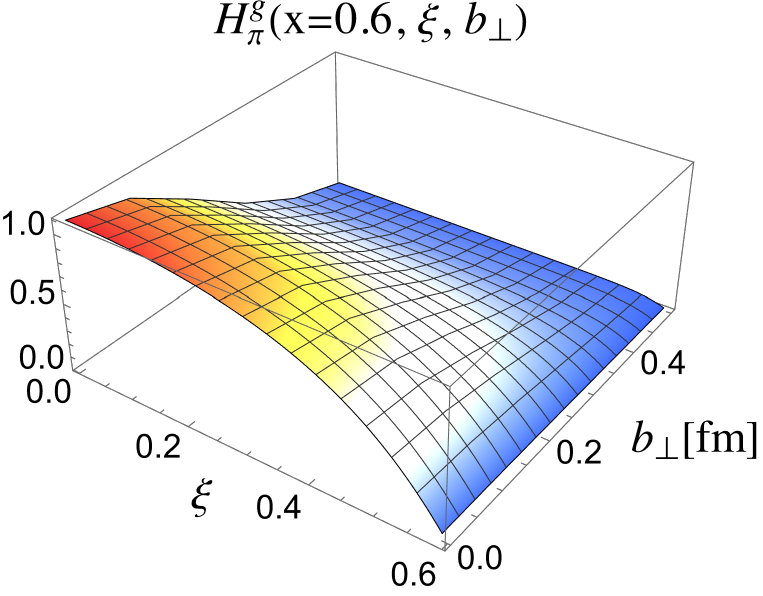}
         \caption{}
         \label{fig:gpd-position-fixed-t}
     \end{subfigure}
\caption{The gluon GPD of the pion in transverse position space, restricted to the DGLAP region ($\xi < x < 1$), as a function of (a) \( x \) and \( b_\perp \) at fixed skewness, \( \xi = 0.15 \), and (b) \( \xi \) and \( b_\perp \) at fixed \( x = 0.6 \).}

        \label{fig:skewed-GPD-position-space}
\end{figure}

The left panel of Fig.~\ref{fig:skewed-GPD-position-space} shows the pion's gluon GPD in the transverse position space, when the pion transfers $15\%$ of the longitudinal momentum to its final state, i.e., at $\xi=0.15$.
The right panel of Fig.~\ref{fig:skewed-GPD-position-space} shows the behavior of the gluon GPD  as a function of skewness \(\xi\) and the transverse impact parameter \({b}_\perp\), for a fixed momentum fraction \(x = 0.6\), meaning the gluon carries 60\% of the pion’s longitudinal momentum. 

As $\xi$ increases, the overall magnitude of the gluon GPD decreases uniformly across all ${b}_\perp$, while its shape in transverse space stays essentially unchanged. This behavior reflects the fact that a larger longitudinal momentum transfer suppresses the GPD amplitude without altering its spatial extent. As \(\xi = x\), the distribution vanishes, as the active gluon in the initial state carries the entire longitudinal momentum and no momentum remains to be transferred. This behavior of the pion’s gluon GPD in transverse position space is broadly consistent with trends observed in gluon GPDs~\cite{Lin:2023ezw,Chakrabarti:2024hwx}, as well as quark GPDs in both the proton~\cite{Mondal:2015uha} and the pion~\cite{Kaur:2018ewq}.

\subsection{Gluon distributions in boost-invariant longitudinal position space}


The longitudinal momentum transfer $\xi P^+$ is Fourier conjugate to the longitudinal distance $\tfrac{1}{2}b^-$, making $\xi$ conjugate to the boost-invariant longitudinal impact parameter $\sigma = \tfrac{1}{2}b^- P^+$. Fourier transforming the GPDs with respect to $\xi$ yields distributions in the boost-invariant longitudinal coordinate $\sigma$, where the DVCS amplitude exhibits a diffraction pattern~\cite{Brodsky:2006in,Brodsky:2006ku}, reminiscent of optical diffraction. This makes the study of GPDs in longitudinal impact-parameter space particularly compelling.

The GPDs in longitudinal impact-parameter space are defined through a Fourier transform with respect to the skewness variable $\xi$:
\begin{equation}
H(x, \sigma, t) = \int_{0}^{\xi_{\rm f}} \frac{{\rm d}\xi}{2\pi} \, e^{i \xi \sigma} H(x, \xi, t),
\end{equation}
where 
%
the upper integration limit, denoted as \(\xi_{\text{f}}\), plays a role analogous to the slit width in diffraction, setting the condition necessary for the appearance of the diffraction pattern. Since the region of interest is \(\xi < x < 1\), the upper limit of the integration is given by \(\xi_{\rm f} = \min(x,\, \xi_{\text{max}})\). For a fixed value of \(-t\), the maximum value of \(\xi\) is determined by~\cite{Brodsky:2000xy}:
\begin{align}
\xi_{\text{max}} = \frac{1}{\sqrt{1 + \frac{4M^2}{-t}}}.
\end{align}

\begin{figure}[hbt!]
     \centering
     \begin{subfigure}[b]{0.48\textwidth}
         \centering
         \includegraphics[width=\textwidth]{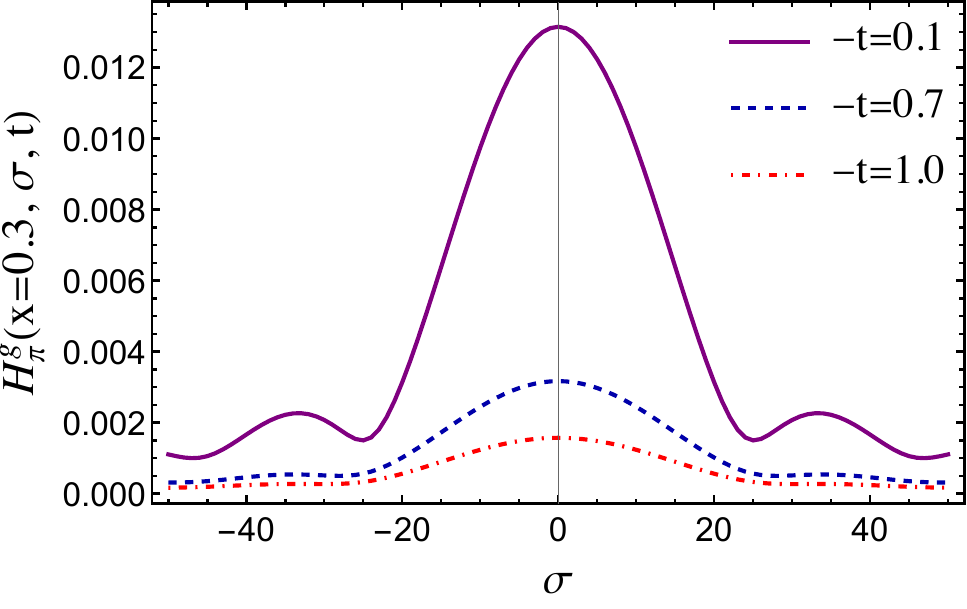}
         \caption{}
         \label{fig:gpd-longi-fixed-zeta}
     \end{subfigure}
    \hfill
     \begin{subfigure}[b]{0.48\textwidth}
         \centering
         \includegraphics[width=\textwidth]{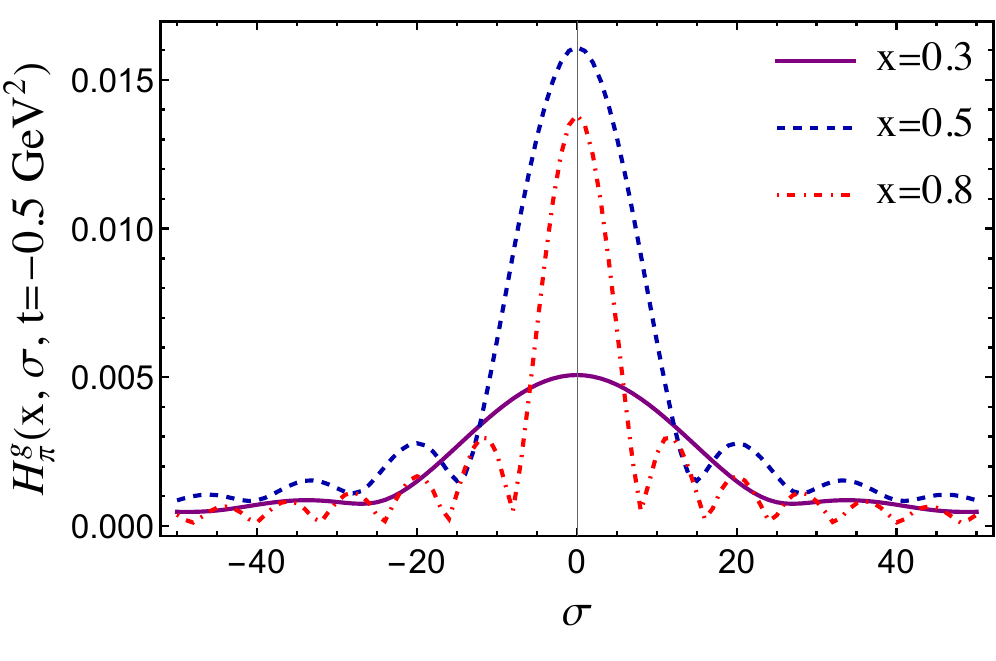}
         \caption{}
         \label{fig:gpd-longi-fixed-t}
     \end{subfigure}
             \caption{
             The gluon GPDs of the pion in longitudinal position space, restricted to the DGLAP region ($\xi < x < 1$), as a function of $\sigma$ (a) at different values of $-t$ [in GeV$^2$], and (b) at different values of $x$.}
        \label{fig:GPD-longi-position-space}
\end{figure}

We present the gluon GPD in longitudinal position space, \( H(x, \sigma, t) \), as a function of the longitudinal impact parameter \( \sigma \), at fixed \( x = 0.3 \) and for different values of momentum transfer squared, \( -t \) in the left panel of  Fig.~\ref{fig:GPD-longi-position-space}. The distribution exhibits a diffraction-like pattern resulting from the Fourier transform with respect to the skewness variable \( \xi \). As \( -t \) increases, the amplitude of the oscillations in \( \sigma \) gradually decreases, indicating a suppression of the distribution. This diminishing pattern reflects reduced coherence between the initial and final states at higher momentum transfers, while the overall shape of the distribution remains largely unchanged. The observation suggests that gluonic contributions to the longitudinal structure become less prominent as the momentum transfer grows.

We also present the gluon GPDs in longitudinal position space, \( H(x, \sigma, t) \), as a function of the longitudinal impact parameter \( \sigma \), for a fixed value of momentum transfer squared \( t = -0.5~\text{GeV}^2 \) and at different values of the longitudinal momentum fraction \( x \) in the right panel of Fig.~\ref{fig:GPD-longi-position-space}. As \( x \) increases, the diffraction pattern becomes more pronounced, with sharper peaks and a more localized distribution in \( \sigma \). This indicates that gluons carrying a larger fraction of the pion’s longitudinal momentum are more tightly localized in longitudinal position space. Conversely, for smaller \( x \), the distribution appears broader, reflecting a more delocalized gluonic structure. This behavior highlights the strong correlation between the momentum fraction and the spatial location of the partons encoded in the GPDs. Note that similar features have also been observed in the gluon GPDs of the proton~\cite{Zhang:2025nll}, as well as in DVCS amplitudes, coordinate-space parton distributions, and Wigner distributions across a range of models~\cite{Brodsky:2006in,Brodsky:2006ku,Miller:2019ysh,Chakrabarti:2008mw,Manohar:2010zm,Kumar:2015fta,Mondal:2015uha,Chakrabarti:2015ama,Mondal:2017wbf,Kaur:2018ewq,Maji:2022tog,Ojha:2022fls}.

\begin{figure}[hbt!]
     \centering
     \begin{subfigure}[b]{0.49\textwidth}
         \centering
         \includegraphics[width=\textwidth]{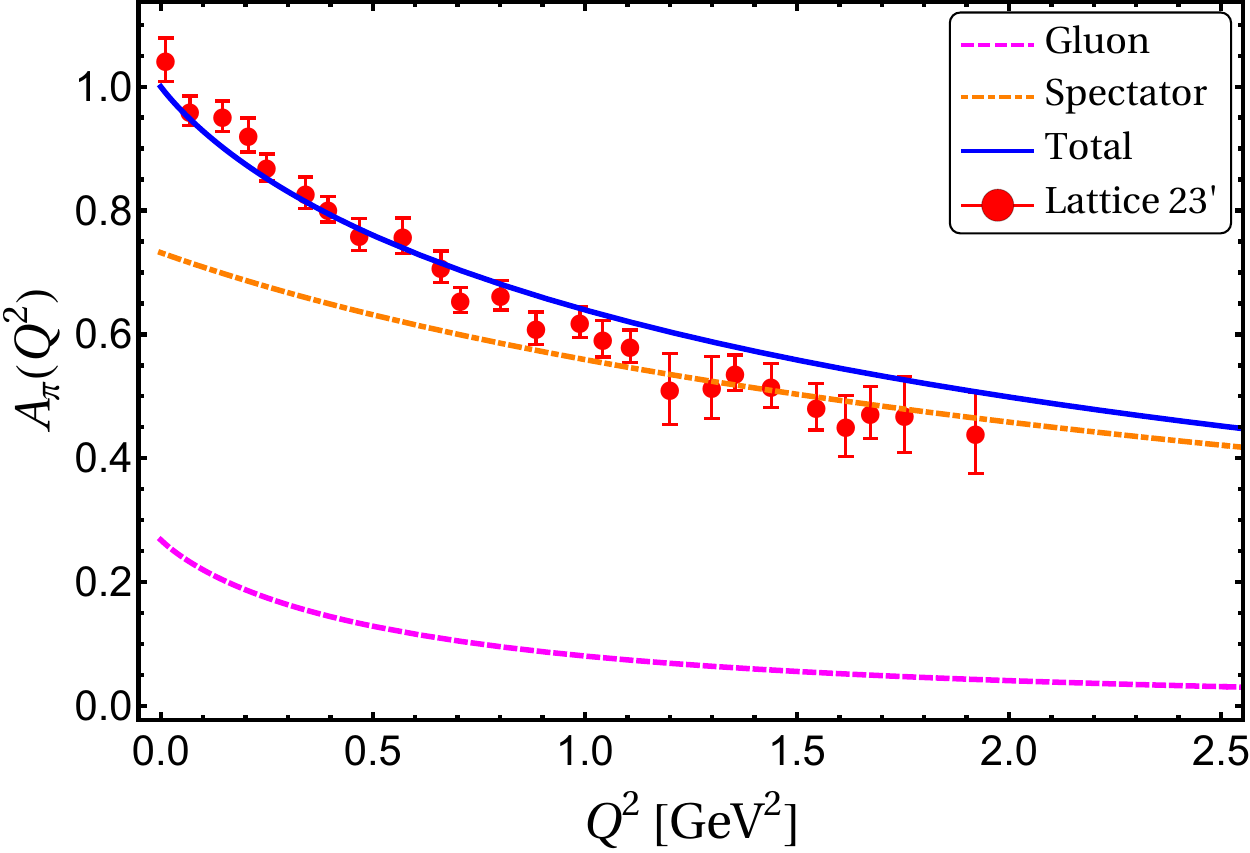}
         \caption{}
         \label{fig:GFF}
     \end{subfigure}
     \begin{subfigure}[b]{0.49\textwidth}
         \centering
         \includegraphics[width=\textwidth]{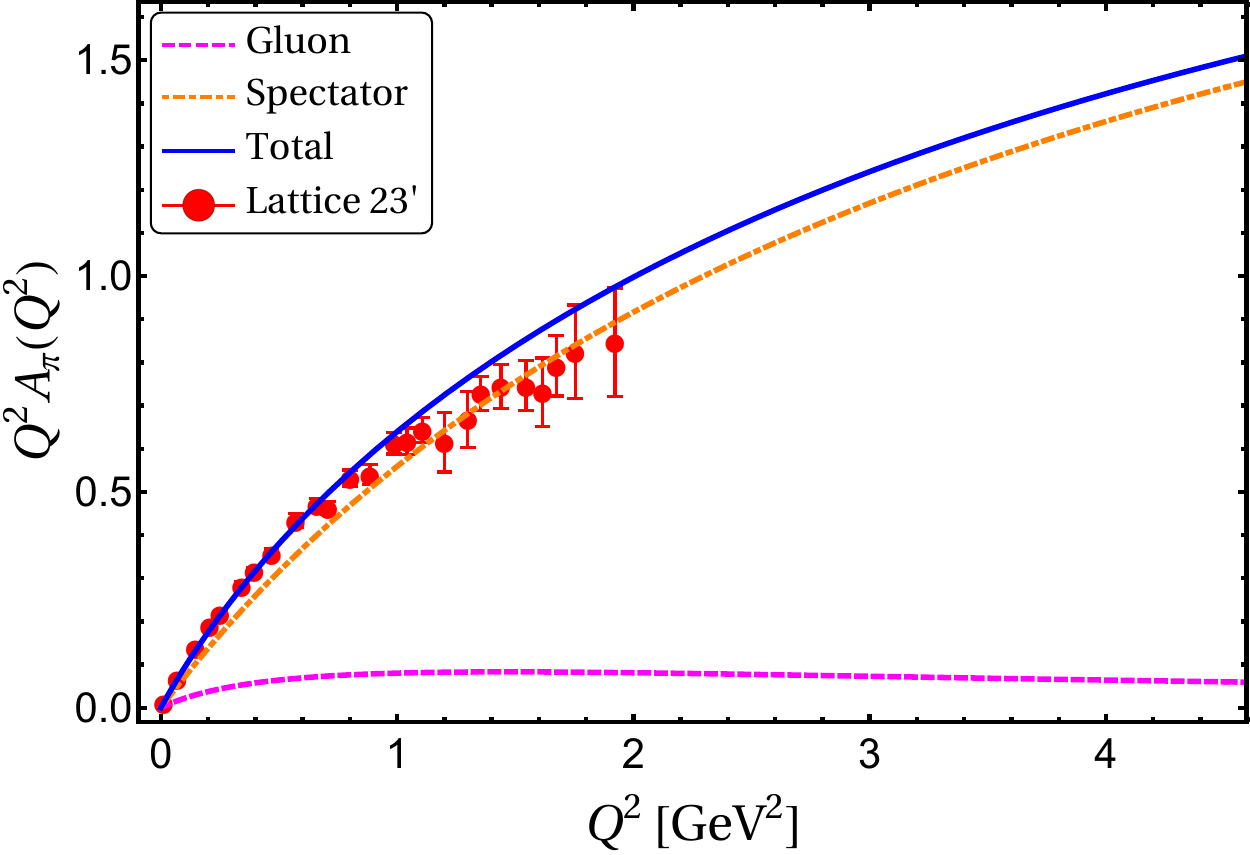}
         \caption{}
         \label{fig:Q2GFF}
     \end{subfigure}
           \caption{Our prediction for the GFF of the pion, showing separate contributions from the gluon (black dashed line) and the spectator (orange dot-dashed line) at the model scale: (a) \( A(Q^2) \), and (b) \( Q^2 A(Q^2) \). The total GFF (blue solid line) is compared with the lattice QCD results~\cite{Hackett:2023nkr}.}
        \label{fig:GFFA}
\end{figure}
 \begin{figure}[hbt!]
     \centering
     \begin{subfigure}[b]{0.47\textwidth}
         \centering
         \includegraphics[width=\textwidth]{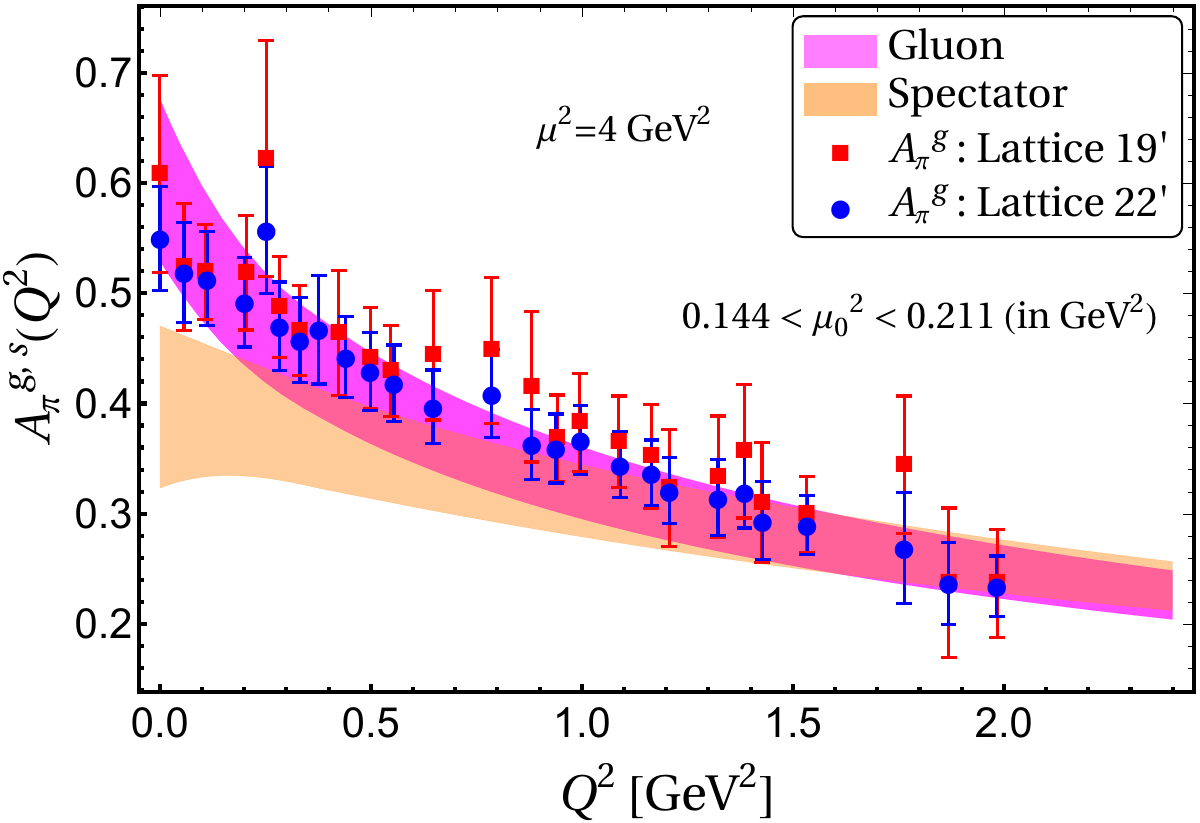}
         \caption{}
         \label{fig:GFF-evolve_scale}
     \end{subfigure}
     \begin{subfigure}[b]{0.47\textwidth}
         \centering
         \includegraphics[width=\textwidth]{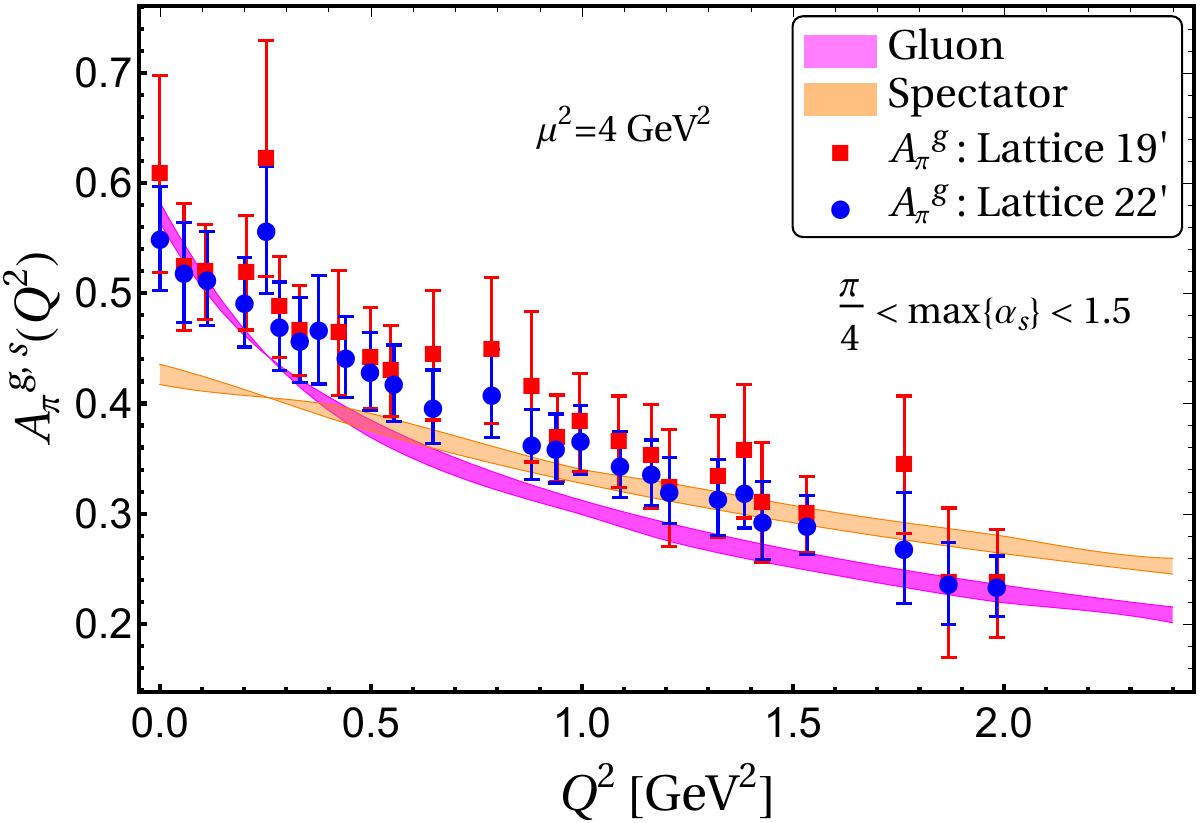}
         \caption{}
         \label{fig:GFF-evolve_as}
     \end{subfigure}
     \begin{subfigure}[b]{0.47\textwidth}
         \centering
         \includegraphics[width=\textwidth]{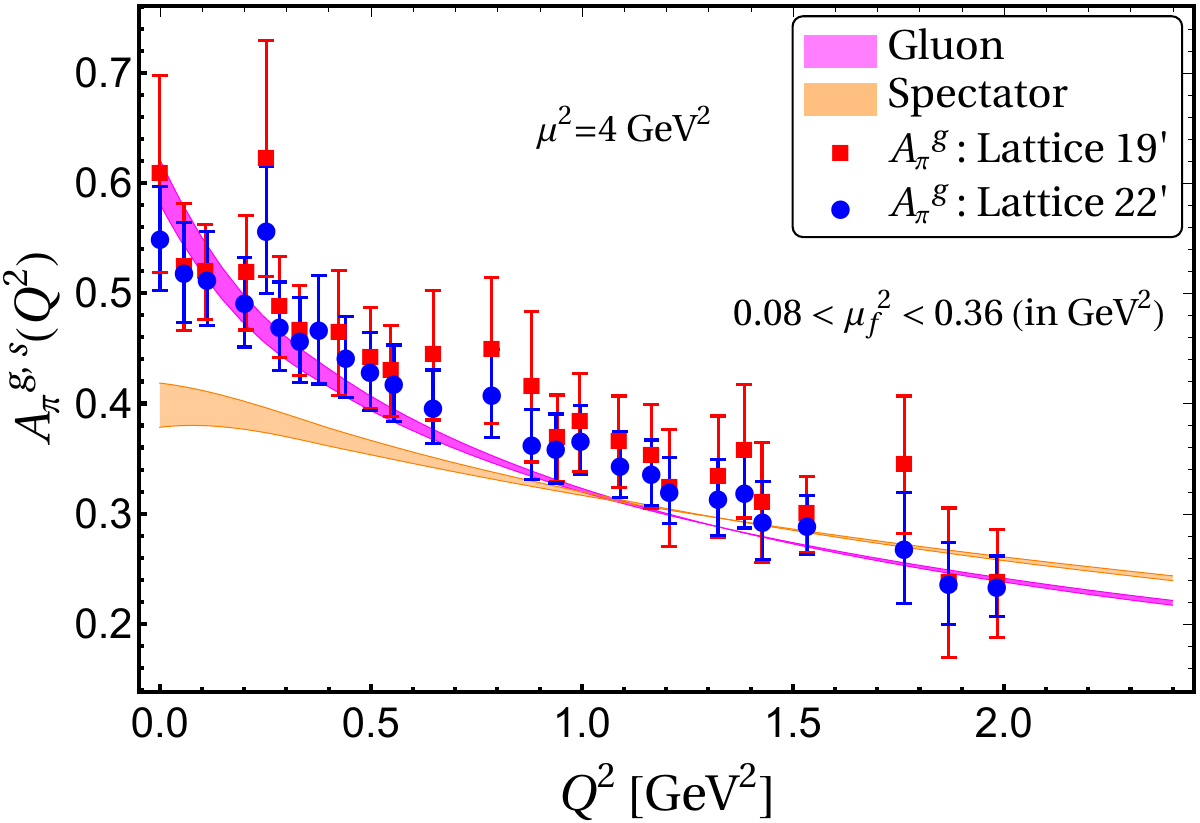}
         \caption{}
         \label{fig:GFF-evolve_muf}
     \end{subfigure}
             \caption{
Gluon (magenta bands) and spectator (orange bands) contributions to the pion GFF $A_\pi(Q^2)$ at the scale $4~\mathrm{GeV}^2$ evolved via DGLAP with different model scale and  IR-freezing prescriptions: (a) varying the model scale $\mu_0^2$ in $0.144 < \mu_0^2 < 0.211~\mathrm{GeV}^2$;
(b) varying $\alpha_s^{\mathrm{max}}$ in the range $\frac{\pi}{4} < \alpha_s^{\mathrm{max}} < 1.5$; 
(c) varying the freeze scale $\mu_f^2$ in $0.08 < \mu_f^2 < 0.36~\mathrm{GeV}^2$. Our results for the gluon GFF, $A_\pi^g (Q^2)$, are compared with the lattice QCD simulations (solid square~\cite{Shanahan:2018pib} and  solid circle~\cite{Hackett:2023nkr}).
}

        \label{fig:GFFas}
\end{figure}

\subsection{Pion gravitational form factor}
The GFFs are connected to the matrix elements of the energy-momentum tensor, while the moment of GPDs also yields the GFFs. 
The  moment of the $H^g_{\pi}(x,\,0,\,t)$ gives the GFF $A^g_\pi(-t=\Delta_\perp^2)$. The LFWF representation of the pion GFF $A(Q^2)$ can be written as 
\begin{align}
A_\pi(Q^2) = &\sum_{\lambda_g, \lambda_s} \int \frac{{\rm d}x\, {\rm d}^2 \vec{k}_\perp}{16\pi^3} \, \psi_{\lambda_g, \lambda_s}\left(x, \vec{k}_\perp\right) \Big[ (1 - x)\, \psi_{\lambda_g, \lambda_s}^*\left(x, \vec{k}_\perp - x\vec{\Delta}_\perp\right)\nonumber\\
&+ x\, \psi_{\lambda_g, \lambda_s}^*\left(x, \vec{k}_\perp + (1 - x)\vec{\Delta}_\perp\right) \Big],
\label{eq:A_WF}
\end{align}
where $Q^2=\Delta_\perp^2$. Figure~\ref{fig:GFFA} shows the GFF $A(Q^2)$ of the pion, including separate contributions from the gluon and the spectator at the model scale. 

We emphasize that the decomposition into ``gluon" and ``spectator" contributions is a model partition arising from the adopted LFWFs; it does not correspond to a unique, field-theoretic separation of QCD energy-momentum operators. By construction, the normalization of our wave functions ensures
\begin{equation}
A_\pi(0) = A^g_\pi(0) + A^s_\pi(0) = 1,
\end{equation}
preserving momentum conservation at zero momentum transfer. At the model scale, we observe that the spectator contribution strongly dominates over the gluonic contribution. We compare our total result, $A_\pi(Q^2)$, with the latest lattice QCD simulations~\cite{Hackett:2023nkr} and find good agreement. From this result, we extract the pion’s mass (or matter) radius as $r^2_{\rm m} = -6A'(0)/A(0) = (0.46~\mathrm{fm})^2$, which is very close to the corresponding lattice QCD value $r^2_{\rm m} = (0.41~\mathrm{fm})^2$~\cite{Hackett:2023nkr}.

While the total GFFs are renormalization-scheme and scale independent, the individual gluonic and spectator contributions, which can be interpreted as moments of the corresponding GPDs, are scale dependent. We therefore perform the QCD evolution of the pion’s gluon GFF, $A_\pi^g(Q^2)$, to the relevant lattice-QCD scale, $\mu^2 = 4~\mathrm{GeV}^2$, following the same prescriptions used for the gluon PDF. The evolved GFF results for the gluon and spectator contributions are presented in Fig.~\ref{fig:GFFas}. Our results for the gluon GFF are compared with lattice-QCD simulations~\cite{Shanahan:2018pib,Hackett:2023nkr}, showing good agreement between the two.



\section{Transverse momentum-dependent gluon distributions}
As three-dimensional distribution functions, TMDs offer probabilistic insight into finding a parton with longitudinal momentum fraction \( x \) and transverse momentum squared \({k}^2_\perp \). At equal light-front time \( z^+ = 0 \), the unpolarized TMD is defined through a correlator involving two gluon fields evaluated at different spacetime positions~\cite{Mulders:2000sh,Meissner:2007rx},
\begin{align}
f_{1\pi}^g(x, {k}^2_\perp) 
= \frac{1}{x \bar{P}^+} \int \frac{\mathrm{d}z^- \mathrm{d}^2 \vec{ z}_\perp}{(2\pi)^3} \, 
e^{i x\bar{P}^+ z^-}
\langle \pi(P) | F^{+i}\left(-\frac{z}{2}\right) F^{+i}\left(\frac{z}{2}\right) | \pi(P) \rangle\Big|_{z^+ = 0}.
\label{eqn:tmd2}
\end{align}
The gluon TMDs can be expressed in terms of overlaps of LFWFs as follows~\cite{Chakrabarti:2023djs}
\begin{align}
&f_{1\pi}^g(x, {k}^2_\perp) = \frac{1}{16\pi^3}\sum_{\lambda_g,\lambda_s}  \left|\psi^\pi_{\lambda_g,\lambda_s} (x,\vec{k}_\perp)\right|^2.
\label{eqn:tmdwf}
\end{align}

 \begin{figure}[hbt!]    
   \centering
\includegraphics[width=0.5\textwidth]{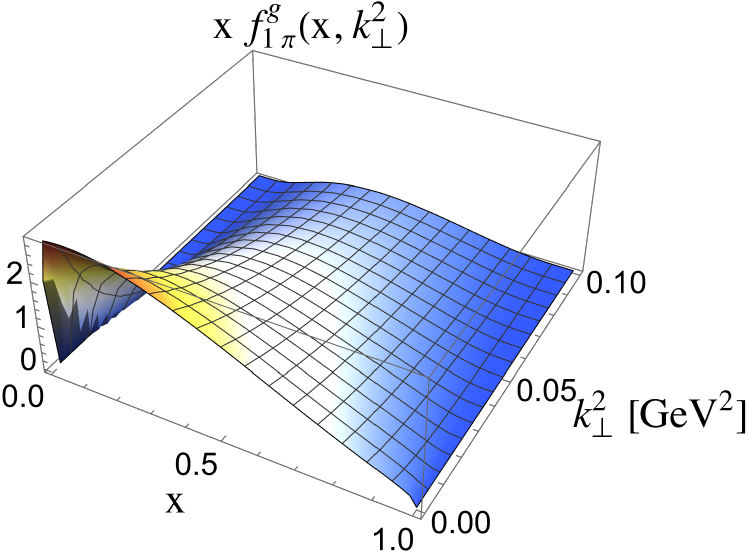}
\caption{The $x$-weighted gluon TMD, $f_{1 \pi}^g(x, k_\perp^2)$, in the pion as a function of $x$ and $k_\perp^2$.}

\label{fig:xTMD}
\end{figure}

In Fig.~\ref{fig:xTMD}, we present the gluon TMD in the pion, weighted by the momentum fraction $x$. This quantity represents the probability density of finding a gluon inside the pion with a given longitudinal momentum fraction $x$ and transverse momentum $\vec{k}_\perp$. The figure reveals key features of the gluonic structure of the pion in momentum space. As expected from the general behavior of gluon distributions in hadrons, the density is largest at small $x$, reflecting the dominant role of soft gluons in the nonperturbative regime of QCD. These low-$x$ gluons are abundant due to gluon radiation and the steep growth of gluon densities at small momentum fractions, a trend commonly observed in both global fits and theoretical models.

In the transverse momentum direction, the distribution exhibits a peak near $\vec{k}_\perp = 0$, indicating that gluons are more likely to be found with low transverse momenta. The density falls off monotonically with increasing $\vec{k}_\perp$, eventually vanishing at large $\vec{k}_\perp$, as expected in a confined system where high-momentum components are suppressed. This behavior is consistent with confinement dynamics and the absence of free gluons in the hadronic spectrum.

The suppression of gluon TMDs at large $x$ and large $\vec{k}_\perp$ reflects the fact that high-momentum gluons are less likely to emerge from a predominantly valence-like structure at the model scale. Furthermore, the shape of the distribution can serve as an input for QCD evolution equations, such as the Collins--Soper--Sterman formalism~\cite{Collins:1984kg}, to make predictions at higher scales relevant for experimental observables in gluon-sensitive processes, such as pion-induced quarkonium production, DY, and dijet production.




\begin{figure}[hbt!]
	\centering
\includegraphics[width=0.5\textwidth]{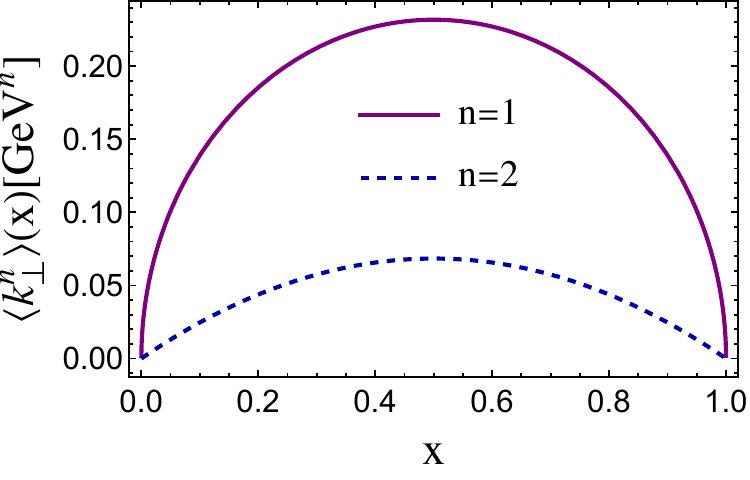}
\caption{The first and second \( k_\perp \)-moments of the gluon TMD \( f_{1 \pi}^g(x, k_\perp^2) \) in the pion, shown as functions of \( x \).}
\label{k-moment}
\end{figure}

Further, we compute the first and second transverse momentum moments of the pion's gluon TMD \( f_{1\pi}^g \), which characterize the average and the spread of the gluon's transverse motion inside the pion. These moments are defined as~\cite{Lorce:2016ugb}
\begin{equation}
    \langle k_\perp^n \rangle^g_\pi(x) = \frac{\int \mathrm{d}^2 \vec{k}_\perp \, k_\perp^n f_{1\pi}^g(x, k_\perp^2)}{\int \mathrm{d}^2 \vec{k}_\perp \, f_{1\pi}^g(x, k_\perp^2)},
\end{equation}
where \( n \) denotes the order of the moment.
Figure~\ref{k-moment} shows the \( x \)-dependence of the first (\( n=1 \)) and second (\( n=2 \)) transverse momentum moments. We observe that both \( \langle k_\perp^n \rangle^g_\pi \) display a strong dependence on the gluon's longitudinal momentum fraction \( x \), peaking around \( x = 0.5 \), and exhibiting approximate symmetry about this point. This behavior reflects the underlying symmetry of the pion’s internal structure in our model. The peak at intermediate \( x \) suggests that gluons carrying a moderate fraction of the pion's momentum are more likely to possess significant transverse motion. These transverse momentum moments offer valuable insight into the interplay between longitudinal and transverse dynamics of gluons and are essential inputs for TMD phenomenology and small-$x$ physics.

In addition to the $x$-dependent moments, we also compute the first and second transverse momentum moments of the gluon TMD integrated over the full range of the gluon's longitudinal momentum fraction. These $x$-integrated moments quantify the overall transverse motion of gluons inside the pion and are defined as~\cite{Lorce:2016ugb}
\begin{equation}
    \langle k_\perp^n \rangle^g_\pi = \frac{\int_0^1 \mathrm{d}x \int \mathrm{d}^2 \vec{k}_\perp \, k_\perp^n f_{1\pi}^g(x, k_\perp^2)}{\int_0^1 \mathrm{d}x \int \mathrm{d}^2 \vec{k}_\perp \, f_{1\pi}^g(x, k_\perp^2)}.
\end{equation}
These global observables offer a complementary perspective to the $x$-dependent moments, providing an averaged measure of gluon transverse dynamics across all momentum fractions. Numerically, we find the average transverse momentum and its square to be
\[
\langle k_\perp \rangle^g_\pi = 0.175~\text{GeV}, \qquad \langle k_\perp^2 \rangle^g_\pi = 0.043~\text{GeV}^2,
\]
indicating that gluons inside the pion typically carry modest transverse momentum at the model scale. These integrated moments are also useful benchmarks for comparing different theoretical models and lattice extractions of TMDs.

\section{Conclusions}
We have developed a light-front framework to study the gluonic structure of the pion by explicitly incorporating gluon degrees of freedom. In this model, high-energy interactions are mediated by an active gluon, while the rest of the pion is described as a spectator system with an effective mass.
The mass spectrum and light-front wave functions (LFWFs) of the pion are simultaneously determined by solving the light-front holographic equation in the chiral limit for transverse dynamics and the 't~Hooft equation for longitudinal confinement. The resulting mass spectrum is consistent with experimental data, validating the predictive power of our framework.

Using the LFWFs, we computed a range of gluon distributions in the pion, including the parton distribution functions (PDFs), generalized parton distributions (GPDs) at nonzero skewness in the DGLAP region, and transverse momentum-dependent distributions (TMDs). 
Our study focuses on the DGLAP region of the pion’s gluon GPD, with the ERBL domain not explicitly constructed. As a result, exact polynomiality of Mellin moments is not satisfied. A minimal extension using a double-distribution model with a D-term illustrates that the DGLAP input is consistent and provides a baseline for future work including the full ERBL region.
The extracted gluon PDF shows good agreement with global fits such as JAM and xFitter, while the gravitational form factor $A(t)$ is found to be in close agreement with recent lattice QCD results. These observables, particularly the GPDs and TMDs, provide essential insights into the three-dimensional gluonic structure of the pion and serve as predictions for upcoming experiments at facilities such as the Electron-Ion Collider in USA and China, Jefferson Lab etc. These findings highlight the effectiveness of our approach in providing a coherent, relativistic, and QCD-based description of pion structure.

\section*{Acknowledgement}
We thank Tobias Frederico for insightful and fruitful discussions. SK is supported by Research Fund for International Young
Scientists, Grant No. 12250410251, from the National Natural Science Foundation of China (NSFC), and China Postdoctoral Science
Foundation (CPSF), Grant No. E339951SR0. CM is supported by new faculty start up funding by the Institute of Modern Physics, Chinese Academy of Sciences, Grant No. E129952YR0.  CM also thanks the Chinese Academy of Sciences Presidents International Fellowship Initiative for support via Grants No. 2021PM0023.

\appendix
\section{Forward limits, normalization, endpoint behavior, and polynomiality checks}\label{appendix}

In this Appendix, we explicitly demonstrate the forward limits, normalization, endpoint behavior, and polynomiality properties of the pion’s gluon distributions. All derivations are based on the same single LFWF employed throughout this work.


The gluon GPD is written as an overlap of the pion LFWFs:
\begin{equation}
H^{g}_\pi(x,\,\xi,\,t) =
\sum_{\lambda_g,\lambda_s} \int \left[{\rm d}\mathcal{X} \,{\rm d}\mathcal{K}_\perp\right]\, 
\psi_{\lambda_g,\lambda_s}^{\pi*}(x^{\prime\prime}, \vec{k}_\perp^{\prime\prime}) \psi_{\lambda_g,\lambda_s}^{\pi}(x^{\prime}, \vec{k}_\perp^{\prime}),
\end{equation}
where the integration measure is defined as
\begin{equation}
\left[{\rm d}\mathcal{X} \,{\rm d}\mathcal{K}_\perp\right] \equiv
\frac{\sqrt{x^2-\xi^2}}{\sqrt{1-\xi^2}}
\prod_{i=1}^{2} \left[ \frac{{\rm d}x_i \, {\rm d}^2 \vec{k}_{i\perp}}{16\pi^3} \right]
\delta(x-x_1) \, 16 \pi^3 \delta \Big(1-\sum_{i=1}^{2} x_i \Big) \delta^2 \Big(\sum_{i=1}^{2} \vec{k}_{i\perp} \Big).
\end{equation}

\subsection{ Forward limit: $H_\pi^g(x,0,0)=x f^g_\pi(x)$}

In the forward limit, $\xi = 0$ and $t = 0$, the shifted momenta satisfy $x' = x'' = x$ and $\vec{k}_\perp' = \vec{k}_\perp'' = \vec{k}_\perp$. Then the GPD reduces to
\begin{align}
H_\pi^g(x,0,0) &= x\sum_{\lambda_g,\lambda_s} \int \frac{{\rm d}^2 \mathbf{k}_\perp}{16 \pi^3} 
|\psi_{\lambda_g,\lambda_s}^\pi(x, \mathbf{k}_\perp)|^2 \nonumber\\
&= x \, f^g_\pi(x),
\end{align}
where $f^g_\pi(x)$ is the unpolarized gluon PDF at the model scale.

This relation demonstrates that the forward limit of the GPD reproduces the PDF, as expected from general principles.

\subsection{Endpoint behavior and consistency tests}

We have performed several checks to verify that our constructed pion gluon GPDs satisfy
basic theoretical constraints at the model scale.

\paragraph{(i) Support}:
By construction, the gluon GPD $H_\pi^g(x,\xi,t)$ and its forward limit $f^g_\pi(x)$ 
are defined through convolution integrals over internal light-front momentum fractions, 
ensuring support in the physical region $0 < x < 1$. 
This property has been verified numerically for all $t$ and $\xi$, as illustrated in Figs.~\ref{gpd:zero} and \ref{fig:GPDas}.

\begin{figure}[hbt!]
	\centering
\includegraphics[width=0.5\textwidth]{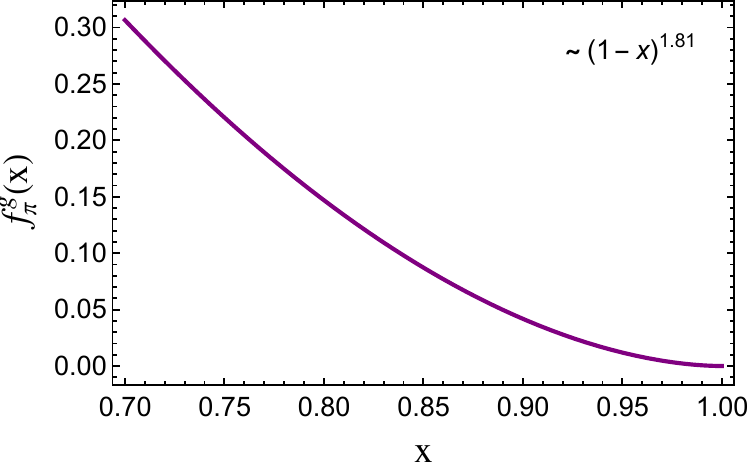}
\caption{
Large-$x$ behavior of our pion's gluon PDF at the model scale.
}
\label{large-x}
\end{figure}

\paragraph{(ii) Large-$x$ behavior}:
At large $x$, the gluon PDF follows a falloff consistent with the perturbative QCD 
counting-rule expectation for a spin-0 hadron,
\begin{equation}
f^g_\pi(x) \sim (1 - x)^{\beta_g}, 
\qquad \beta_g \simeq 2n_s - 1 = 3,
\label{eq:large_x_falloff}
\end{equation}
where $n_s = 2$ is the number of spectator quarks within $q\bar{q}g$ Fock component. 
A fit to our model result, shown in Fig.~\ref{large-x}, yields $\beta_g \approx 1.81$, which is lower compared to that in Eq.~(\ref{eq:large_x_falloff}). The smaller $\beta_g$ value is expected, as the $q\bar{q}$ pair is treated as an effective diquark spectator in our approach.

\paragraph{(iii) Small-$x$ behavior}:
No explicit Regge-like input is introduced in the model. 
The moderate rise of $x f^g_\pi(x,\mu^2)$ toward small $x$ (Figs.~\ref{fig:PDF} and \ref{fig:PDFas}) emerges dynamically 
through DGLAP evolution from the valence-like input at $\mu_0^2$. 
Thus, the small-$x$ enhancement is entirely generated by QCD evolution.

\begin{figure}[hbt!]
	\centering
\includegraphics[width=0.5\textwidth]{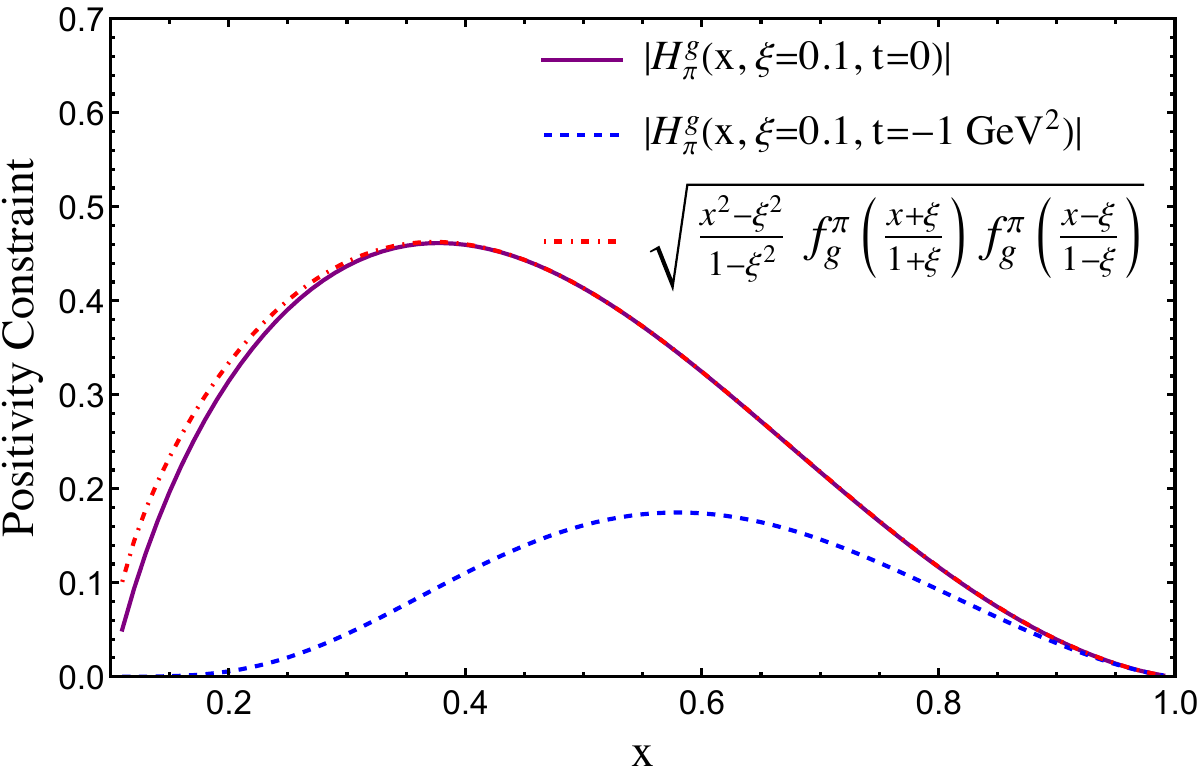}
\caption{
The positivity bounds of the pion's gluon
GPDs versus $x$ at fixed for different $-t$ values.
}
\label{fig:positivity}
\end{figure}

\paragraph{(iv) Positivity bounds}:
We have verified that the gluon GPD satisfies the standard positivity inequality 
for a spin-0 target in the DGLAP region\cite{Diehl:2000xz},
\begin{equation}
|H_\pi^g(x,\xi,t)| 
\le 
\sqrt{\frac{x^2-\xi^2}{1-\xi^2}
f^g_\pi\!\left(\frac{x+\xi}{1+\xi}\right)
\, f^g_\pi\!\left(\frac{x-\xi}{1-\xi}\right)
},
\label{eq:positivity}
\end{equation}
within numerical accuracy. 
This relation provides a nontrivial cross-check of the internal consistency of our GPD construction.
Representative results are shown in Fig.~\ref{fig:positivity}.

\begin{figure}[hbt!]
	\centering
\includegraphics[width=0.5\textwidth]{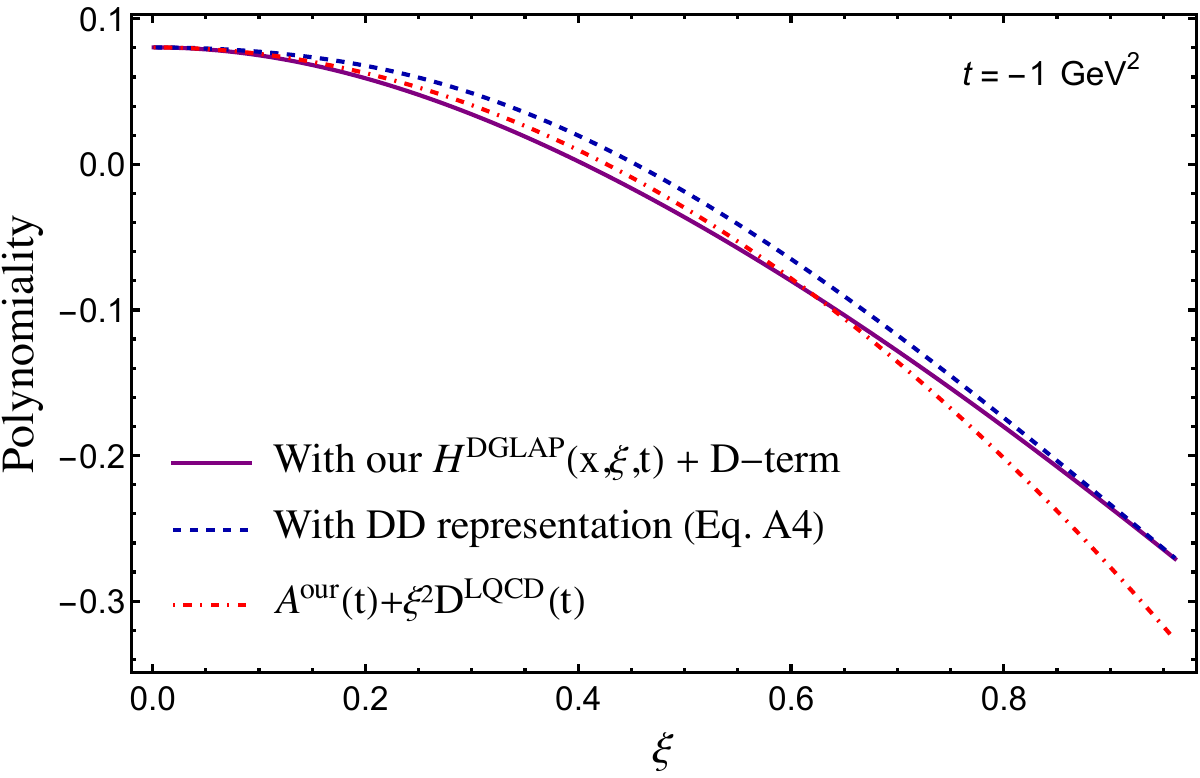}
\caption{
Verification of the polynomiality property of the pion's gluon GPD. 
The first Mellin moment of the gluon GPD follows a second-order polynomial in $\xi$, 
of the form $A_\pi^g(t) + \xi^2 D_\pi^g(t)$, see Eq.~\eqref{eq:polynomialityn1}. 
The solid magenta line shows the result obtained using our DGLAP-region GPD embedded 
into the ERBL region through a modeled $D$-term. 
The blue dashed line represents the result based on the standard double-distribution (DD) representation. 
The red dot-dashed line corresponds to the reference polynomial 
$A_\pi^g(t) + \xi^2 D_\pi^g(t)$, where $A_\pi^g(t)$ is calculated within our model 
and $D_\pi^g(t)$ is adopted from lattice QCD~\cite{Hackett:2023nkr}.
}
\label{polynomiality}
\end{figure}

\subsection{Minimal double-distribution completion and polynomiality check}\label{DD}

To illustrate the effect of the missing ERBL contribution and verify the consistency of our DGLAP-region result, we embed the obtained gluon GPD into a minimal double-distribution (DD) ansatz supplemented by a phenomenological $D$-term. The gluon GPD can be expressed in the standard DD representation as~\cite{Diehl:2003ny}
\begin{equation}
H_\pi^g(x,\xi,t) = \int_{0}^{1} \! dy \int_{-1+y}^{1-y} \! dz \,
\delta(x - y - \xi z)\, F_\pi^g(y,z,t)
+ \theta(|x|<\xi)\, |\xi|\, \mathcal{D}_\pi^g\!\left(\frac{x}{\xi},t\right),
\label{eq:DD_representation}
\end{equation}
where $F_\pi^g(y,z,t)$ denotes the double distribution and $\mathcal{D}_\pi^g(x/\xi,t)$ is the so-called $D$-term, which contributes solely in the ERBL region ($|x|<\xi$).

For a minimal completion, we model the double distribution as a factorized form
\begin{equation}
F_\pi^g(y,z,t) = g_\pi(y,t)\, \frac{3}{4}\, \frac{(1-y)^2 - z^2}{(1-y)^3},
\label{eq:DD_model}
\end{equation}
where $g_\pi(y,t)$ is the gluon GPD in the DGLAP domain obtained from our calculation. The simple profile function in Eq.~(\ref{eq:DD_model}) ensures smooth matching between the DGLAP and ERBL regions. The $D$-term is modeled as
\begin{equation}
\mathcal{D}_\pi^g\!\left(\frac{x}{\xi},t\right) = d_0(t)\,\left(1 - \frac{x^2}{\xi^2}\right)^2,
\label{eq:Dterm_model}
\end{equation}
with $d_0(t)$ adjusted such that the  Mellin moments satisfy the polynomiality condition. Here we employ the dipole form of $d_0(t)$ adopted from lattice QCD~\cite{Hackett:2023nkr}:
\begin{align}
    d_0(t)=\frac{5}{4} D_\pi^g=\frac{5}{4}\frac{\alpha}{(1-\frac{t}{m^2})^2}\,,
\end{align}
with $\alpha=-1.2$ and $m=1.24$ GeV.

The corresponding sum rule of the GPD reads~\cite{Diehl:2003ny}
\begin{equation}
\int_{0}^{1} \! dx \, x^{\,n-1} \, H_\pi^{g}(x,\xi,t)
= \sum_{\substack{i=0 \\ \text{even}}}^{n}
(2\xi)^{i} \, A_{n+1,i}^{g}(t)
+ \text{mod}(n,2)\,(2\xi)^{n+1} \, C_{n+1}^{g}(t).
\label{eq:polynomiality}
\end{equation}
For $n=1$, the above equation leads to
\begin{equation}
\int_{0}^{1} \! dx  \, H_\pi^{g}(x,\xi,t)
=  A_{\pi}^{g}(t)
+ \xi^{2} \, D_{\pi}^{g}(t),
\label{eq:polynomialityn1}
\end{equation}
where $4C_2^g(t)=D_\pi^g(t)$.

Using this construction, we verified that the first moment of $H_\pi^g(x,\xi,t)$ follows the expected $\xi$-dependence within numerical precision, as shown in Fig.~\ref{polynomiality}, confirming that our DGLAP input is consistent with a well-behaved DD completion.

\subsection{Transverse-momentum integration of the TMD}

The unpolarized gluon TMD is defined as
\begin{equation}
f_{1\pi}^g(x, {k}_\perp^2) = \sum_{\lambda_g,\lambda_s} \frac{1}{16\pi^3} |\psi_{\lambda_g,\lambda_s}^\pi(x, \mathbf{k}_\perp)|^2.
\end{equation}

Integration over $\mathbf{k}_\perp$ reproduces the PDF:
\begin{align}
\int {\rm d}^2 {k}_\perp \, f_1^g(x, \mathbf{k}_\perp^2) 
&= \sum_{\lambda_g,\lambda_s} \int \frac{{\rm d}^2 \mathbf{k}_\perp}{16\pi^3} |\psi_{\lambda_g,\lambda_s}^\pi(x, \mathbf{k}_\perp)|^2 \nonumber\\
&= f^g_\pi(x).
\end{align}

\subsection{Gravitational form factor normalization: $A_\pi(0)=1$}

The gluon contribution to the gravitational form factor is given by the Drell–Yan-frame overlap:
\begin{align}
A_\pi(Q^2) = &\sum_{\lambda_g, \lambda_s} \int \frac{{\rm d}x\, {\rm d}^2 \vec{k}_\perp}{16\pi^3} \, \psi_{\lambda_g, \lambda_s}\left(x, \vec{k}_\perp\right) \Big[ (1 - x)\, \psi_{\lambda_g, \lambda_s}^*\left(x, \vec{k}_\perp - x\vec{\Delta}_\perp\right)\nonumber\\
&+ x\, \psi_{\lambda_g, \lambda_s}^*\left(x, \vec{k}_\perp + (1 - x)\vec{\Delta}_\perp\right) \Big],
\end{align}
where $\vec{\Delta}_\perp$ is the transverse momentum transfer and $t=-{\Delta}_\perp^2$. The weighting factors \((1-x)\) and \(x\) account for the momentum fractions carried by the spectator and active gluon, respectively.

At zero momentum transfer, $\vec{\Delta}_\perp = 0$, we obtain
\begin{align}
A_\pi(0) &= \sum_{\lambda_g,\lambda_s} \int_0^1 dx \int \frac{{\rm d}^2 \mathbf{k}_\perp}{16 \pi^3} 
|\psi_{\lambda_g,\lambda_s}^\pi(x, \mathbf{k}_\perp)|^2 \nonumber\\
&= 1.
\end{align}
 This demonstrates that the LFWF correctly reproduces the momentum sum rule and normalizes the GFF at zero momentum transfer.


\bibliographystyle{apsrev}
\bibliography{ref}

 \end{document}